\documentclass[10pt,a4paper]{article}
\usepackage[utf8]{inputenc}
\usepackage{amsmath}
\usepackage{amsfonts}
\usepackage{amssymb}
\usepackage{MnSymbol,wasysym}
\usepackage{subfigure}
\usepackage{tikz}
\usepackage{url}
\usepackage{graphics}
\usepackage{tkz-euclide}
\usepackage[toc,page]{appendix}
\usetikzlibrary{decorations.markings,arrows}
\usetikzlibrary{arrows,shapes,backgrounds}
\usetikzlibrary{decorations.pathreplacing,decorations.markings}
\usetikzlibrary{calc, trees, positioning, arrows, shapes, shapes.multipart, shadows, matrix, decorations.pathreplacing, decorations.pathmorphing}
\usetikzlibrary{shapes.misc}
\usepackage{youngtab}
\usepackage{rotating}
\usepackage{graphicx}
\usepackage{slashed}
\usepackage{a4wide}
\usepackage{appendix}
\usepackage{hyperref}
\usepackage{bookmark}
\usepackage{verbatim}
\usepackage{footmisc}
\usepackage[yyyymmdd,hhmmss]{datetime}

\def\makeatletter{\catcode`\@=11}
\makeatletter
\def\mathbox#1{\hbox{$\m@th#1$}}%
\def\math@ccstyles#1#2#3#4#5#6#7{{\leavevmode
      \setbox0\mathbox{#6#7}%
      \setbox2\mathbox{#4#5}%
      \dimen@ #3%
      \baselineskip\z@\lineskiplimit#1\lineskip\z@
      \vbox{\ialign{##\crcr
             \hfil \kern #2\box2 \hfil\crcr
             \noalign{\kern\dimen@}%
             \hfil\box0\hfil\crcr}}}}
\def\mathaccstyles{\math@ccstyles\maxdimen}
\def\maththroughstyles{\math@ccstyles{-\maxdimen}}
\def\unity%
 {\maththroughstyles{.45\ht0}\z@\displaystyle {\mathchar"006C}\displaystyle 1}

\begin{document}
\interfootnotelinepenalty=10000000


\begin{titlepage}



\vskip 2cm

\begin{center}
{\Large \bfseries HWG for Coulomb branch of $3d$ Sicilian theory mirrors}

\vskip 1.2cm

Amihay Hanany\textsuperscript{$\clubsuit$,1},
Alessandro Pini\textsuperscript{$\spadesuit$,2}

\bigskip
\bigskip

\begin{tabular}{c}
\textsuperscript{$\clubsuit$} Theoretical Physics Group, Imperial College London\\
Prince Consort Road, London, SW7 2AZ, UK,\\
\\
\textsuperscript{$\spadesuit$}Department of Physics, Universidad de Oviedo, \\
Avda.~Calvo Sotelo 18, 33007, Oviedo, Spain\\
\\

\end{tabular}

\vskip 1.5cm

\textbf{Abstract}
\end{center}

\medskip
\noindent
Certain star shaped quivers exhibit a pattern of symmetry enhancement on the Coulomb branch of $3d$ $\mathcal{N}=4$ supersymmetric gauge theories. This paper studies a subclass of theories where such global symmetry enhancement occurs through a computation of the Highest Weight Generating Function (HWG) and of the corresponding Hilbert Series (HS), providing a further test of the Coulomb branch formula \cite{Cremonesi:2013lqa}. This special subclass has a feature in which the HWG takes a particularly simple form, as a simple rational function which is either a product of simple poles (termed freely generated) or a simple PE (termed complete intersection). Out of all possible star shaped quivers, this is a particularly simple subclass. The present study motivates a further study of identifying all star shaped quivers for which their HWG is of this simple form.
\bigskip
\vfill

\footnotetext[1]{a.hanany@imperial.ac.uk} \footnotetext[2]{pinialessandro@uniovi.es  }

\end{titlepage}

\setcounter{tocdepth}{1}

\tableofcontents

\title{HWG computation for the Coulomb branch of $3d$ quiver gauge theories}
\author{Alessandro Pini}
\date{\today}

\section{Introduction}


Recently a general formula which allows to count BPS gauge invariant operators for the Coulomb branch of $3d$ $\mathcal{N}=4$ theories has been introduced \cite{Cremonesi:2013lqa}. This result is very remarkable since the structure of the chiral ring associated with the Coulomb branch of $3d$ $\mathcal{N}=4$ is quite involved. This is due to the fact that also monopole operators are present in addition to the classical fields in the Lagrangian. However, as stated above, the so-called \textit{monopole formula} \cite{Cremonesi:2013lqa} allows to describe the Coulomb branch using monopole operators dressed with scalar fields from the vector multiplet. This formula reproduces the \textit{Hilbert Series} (HS) for the Coulomb branch, i.e. the generating function which counts chiral operators present in the theory according to their dimension and other quantum numbers under global symmetries (see e.g. \cite{Benvenuti:2006qr} for an introduction to this topic). This formula can be applied to any gauge theory that is good or ugly in the sense of \cite{Gaiotto:2008ak}. Recently a new technique that simplifies the computation of the HS for ``good" theories has been worked out \cite{Hanany:2016ezz,Hanany:2016pfm}. This novel approach relies on the notion of \textit{Hilbert basis}, that is a sufficient set of monopole operators that generates the chiral ring and whose knowledge completely determine the HS. Moreover the application of the monopole formula led to an expression for the Coulomb branch Hilbert Series of the $T_{\rho}(G)$ theory in terms of the Hall-Littlewood polynomials \cite{Cremonesi:2014kwa} and for the Coulomb branch HS of $T_{\rho}^{\sigma}(G)$ theory in terms of the so called generalized Hall-Littlewood polynomials \cite{Cremonesi:2014uva}. Moreover this formula has been successfully applied also in the context of the mirror of $3d$ Sicilian theories \cite{Benini:2010uu,Cremonesi:2014vla}. These theories arise from the compactification of the $6d$ (2,0) theory with symmetry group $G$ on a circle times a Riemann surface with punctures. As we will review the HS of these theories can be obtained by gluing together different $T_{\rho}(G)$ theories.

Moreover recently it has been developed a new mathematical tool that simplifies the computation of the HS, the so called \textit{Highest Weight Generating function} (HWG) (see \cite{Hanany:2014dia} for an introduction to this topic). This method is based on the highest weight Dynkin labels of the symmetry group that characterizes the theory taken under consideration and it has already been successfully  applied \cite{Hanany:2015hxa}\cite{Hanany:2016gbz}.

In the present paper we move a further step in this direction and we perform the computation of the HWG and of the corresponding HS for the mirror of certain $3d$ Sicilian theories \cite{Benini:2010uu}, which are chosen such that they exhibit a sufficiently large global symmetry. In particular we examine how the HS can be decomposed under representations of the global symmetry group that characterizes these theories.

The present paper is organized as follows. In section \ref{sec:review}, after a short review of the Coulomb branch formula introduced in \cite{Cremonesi:2013lqa} and its application in the context of the $T_{\rho}(G)$ theory \cite{Cremonesi:2014kwa}, we examine how such  formula can be applied for the computation of the HS of the mirror of $3d$ Sicilian theories \cite{Cremonesi:2014uva}. Moreover we also summarize the basic aspects of the computation technique that we employ in the following part of the paper. In section \ref{sec:nilpotent} we review the relation between the Coulomb and Higgs branch of $3d$ $\mathcal{N}=4$ theories and closure of nilpotent orbits \cite{nilpotent,2016arXiv160306105N}. In section \ref{sec:results} we summarize our main results, i.e. the general expressions of the HWG for the theories that have been taken into account. We focus our attention on theories with unitary and orthogonal global symmetry groups
and on the mirror of the $(k)-[2N]$ theory (see \cite{Hanany:2011db,Hanany:1996ie}). 
Then in section \ref{sec:hscomp} we test the previous expressions performing the explicit computation of the HWG and of the Plethystic Logarithm (PLog) for theories with unitary global symmetry group. We examine in detail the cases in which the integer $N$, that characterizes the theory, is equal to $3$ and $4$. While we refer the reader to the appendix \ref{app:quiver} for the analysis performed for higher values of $N$. Then in section \ref{sec:globalort} we test the expression of the HWG for theories with orthogonal global symmetry group. Finally we end up with some conclusions in section \ref{sec:conclusions}. We refer the reader to appendices \ref{app:A} - \ref{app:quiver} for the conventions that have been employed and more technical aspects related to the computations.

\section{The Coulomb branch Hilbert Series for $3d$ $\mathcal{N}=4$ theories}
\label{sec:review}
In all this paper we consider the Coulomb branch of $3d$ $\mathcal{N}=4$ gauge theories. This branch is described by the VEVs of the triplet of scalar fields in the $\mathcal{N}=4$ vector multiplet and by the VEV of the dual photons.  Differently from the Higgs branch the Coulomb branch is affected by quantum corrections and the corresponding chiral ring also involves  monopole operators. A suitable description of the chiral ring on the Coulomb branch has been introduced in \cite{Cremonesi:2013lqa}. As a matter of fact the gauge invariant objects in this branch are monopole operators dressed by a product of certain scalar fields in the vector multiplet. This provided a systematic way to study  the chiral ring of this branch. Moreover an analytic expression of the corresponding generating function, known as the \textit{Hilbert Series} (HS), has been found. This function counts gauge invariant BPS operators that have a non-zero vacuum expectation value along the Coulomb branch. In the following we denote this expression as the \textit{monopole formula}. We review this formula in section \ref{subsec:monopole}. 

Using the monopole formula an analytic expression for the HS of the so called $T_{\rho}(G)$ theories \cite{Gaiotto:2008ak} has been introduced in \cite{Cremonesi:2014kwa}. This expression holds for any classical gauge group $G$  and for any partition $\rho$  related to the GNO dual group $G^{v}$ \cite{1977NuPhB.125....1G}. Moreover it has been shown that the HS can be expressed as a function of the Hall-Littlewood polynomials. In the following we denote this formula as the \textit{Hall-Littlewood formula} and we review it in section \ref{subsec:hall}. The previous result has been generalized in the context of the $T_{\rho}^{\sigma}(G)$ theories \cite{Cremonesi:2014uva}.

Finally the previous computational technique has been applied in the context of the mirror of 3-dimensional Sicilian theories \cite{Cremonesi:2014uva}. The computation of the HS for this class of theories can be performed gluing together the HS for different $T_{\rho}(G)$ theories that share the same global symmetry group. We refer to the corresponding formula as the \textit{gluing formula}. We review it in section \ref{subsec:glue}.
In the following part of this section we summarize the basic computational tools that have been employed in the rest of this article.

\subsection{The monopole formula}
\label{subsec:monopole}
The monopole formula \cite{Cremonesi:2013lqa} allows to count all the BPS gauge invariant operators that can acquire a non-zero VEV along the Coulomb branch, according to their dimensions and other quantum numbers. Using the $\mathcal{N}=2$ formalism the $\mathcal{N}=4$ vector multiplet is decomposed in a $\mathcal{N}=2$ vector multiplet and in $\mathcal{N}=2$ chiral multiplet transforming in the adjoint representation of the gauge group. The Hilbert Series for an ugly or a good theory with gauge group $G$ reads \cite{Cremonesi:2013lqa}
\begin{equation}
\label{eq:monopoleform}
\textrm{HS}_{G}(t,z) \  \ = \sum_{\textrm{\textbf{m}} \ \in \ \Gamma_{G^v}/W_{G^v}}z^{J(\textbf{m})}t^{2\Delta(\textbf{m})}P_{G}(t;\textbf{m}),
\end{equation}
where the sum is taken over the magnetic charges $\textbf{m}$ of the monopole operator $V_{\textrm{\textbf{m}}}$ that, modulo a gauge transformation, belongs to a Weyl Chamber of the weight of lattice $\Gamma_{G^v}$ of the GNO dual group \cite{1977NuPhB.125....1G}. The factor $P_{G}(t,\textrm{\textbf{m}})$ counts operators constructed by the adjoint scalar field $\phi$ in the chiral multiplet. These operators are gauge invariant under the action of the gauge group $H_{\textrm{\textbf{m}}}$ unbroken in the presence of the monopole operator $V_{\textrm{\textbf{m}}}$. This factor is given by
\begin{equation}
P_{G}(t;\textbf{m}) = \prod_{i=1}^{r} \frac{1}{1-t^{2d_{i}(\textbf{m})}},
\end{equation}
where the $d_{i}(\textbf{m})$ are the degrees of the independent Casimir invariants of $H_\textrm{\textbf{m}}$. Finally $\Delta(\textrm{\textbf{m}})$ is the dimension of the monopole operator
\begin{equation}
\Delta(\textbf{m}) = - \sum_{\alpha \in \Delta_{+}(G)} \mid \alpha(\textbf{m}) \mid + \frac{1}{2} \sum_{i=1}^{n}\sum_{\rho_i \in R_i}\mid \rho_i(\textbf{m}) \mid,
\end{equation}
where $\alpha$ are the positive roots of the gauge group $G$ and $\rho_i  \in R_i$ are the weights of the matter field representation $R_i$ under the gauge group. $J(\textbf{m})$ is the topological charge, one per each gauge node in the quiver, of the monopole operator of GNO charges $\textbf{m}$. Finally $z$ is the fugacity of the topological symmetry.

The formula (\ref{eq:monopoleform}) can be generalized to also include background monopole fluxes for a global flavour symmetry $G_{F}$ acting on the matter fields. The corresponding Hilbert Series formula reads \cite{Cremonesi:2014kwa}  
\begin{equation}
\label{eq:monopolegf}
\textrm{HS}_{G,G_{F}}(t,\textbf{m}_{F},z)=\sum_{\textbf{m} \ \in \ \Gamma_{G^{v}}/W_{G^{v}}}t^{2\Delta(\textbf{m},\textbf{m}_{F})}P_{G}(t;\textbf{m})z^{J(\textbf{m})},
\end{equation}
where the sum is taken only over the magnetic fluxes of the gauge group $G$ but depends on the weights $\textbf{m}_F$ of the dual flavor group $G_{F}^{v}$. These weights enter in the formula (\ref{eq:monopolegf}) through the  dimension $\Delta$ of the operators. Moreover, using the global symmetry, we can restrict the possible values of $\textbf{m}_F$ to a Weyl chamber of $G_{F}^{v}$ and take $\textbf{m}_F \in \Gamma_{G_{F}^v}/W_{G_{F}^v}$.

\subsection{The Hall-Littlewood formula}
\label{subsec:hall}
The Hilbert Series formula (\ref{eq:monopolegf}) can be applied in the context of the $T_{\rho}(G)$ theory, leading to the Hall-Littlewood formula \cite{Cremonesi:2014kwa}. A $T_{\rho}(G)$ is specified by a partition $\rho$ and classical gauge group $G$.\footnote{For the purpose of this paper the only relevant  case is $G=SU(N)$. Therefore henceforth we focus only on this specific case.}
The partition $\rho$ of $N$ is given by
\begin{equation}
\rho = (N-N_1,N_1-N_2,...,N_{d-1}-N_{d},N_{d}),
\end{equation}
moreover the corresponding theory is ``good" (in the sense of \cite{Gaiotto:2008ak}) if the partitions satisfies the non-increasing constraint
\begin{equation}
N-N_1 \geq N_1-N_2 \geq N_2 -N_3 \geq ... \geq  N_{d-1} -N_{d} \geq N_{d} > 0.
\end{equation}
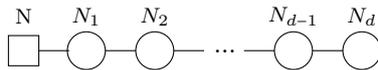
\begin{figure}[h!]
\center{
\begin{tikzpicture}

\draw (0.2,0) -- (0.6,0);
\draw (-0.2,-0.2) rectangle (0.2,0.2);
\draw (0,0.45) node {\footnotesize{N}};

\draw (1.075,0) -- (1.475,0);
\draw (0.825,0) circle (0.25cm);
\draw (0.825,0.45) node {\footnotesize{$N_{1}$}};

\draw (1.975,0) -- (2.375,0);
\draw (1.725,0) circle (0.25cm);
\draw (1.725,0.45) node {\footnotesize{$N_{2}$}};

\draw (2.9,0) -- (3.3,0);
\draw (2.65,0) node {...};

\draw (3.8,0) -- (4.2,0);
\draw (3.55,0) circle (0.25cm);
\draw (3.55,0.45) node {\footnotesize{$N_{d-1}$}};

\draw (4.45,0) circle (0.25cm);
\draw (4.45,0.45) node {\footnotesize{$N_{d}$}};

\end{tikzpicture}
}
\caption{Quiver diagram of the $T_{\rho}(SU(N))$ theory.\label{fig:quivertrho}}
\end{figure}
The quiver diagram for this theory is reported in figure \ref{fig:quivertrho}. The quiver theory can be obtained starting from a brane configurations as proposed in \cite{Hanany:1996ie}. The Hilbert Series for the Coulomb branch of this theory can be expressed in terms of the Hall-Littlewood polynomials as \cite{Cremonesi:2014kwa}
\begin{equation}\begin{split}
\label{eq:hsp}
& \textrm{HS}[T_{\rho}(SU(N))](t;,x_{1},...,x_{d+1},n_{1},...n_{N}) = t^{\delta(\textbf{n})}(1-t^2)^N K_{\rho}(\textbf{x};t)\Psi^{\textbf{n}}_{U(N)}(\textbf{x}t^{t\textbf{w}_{\rho}};t), 
\end{split}\end{equation}
where $n_1, n_2, ..., n_N$ are the background GNO charges for the $U(N)$ group, with
\begin{equation}
n_1 \geq n_2 \geq ... n_{N-1} \geq n_N \geq 0,
\end{equation}
and the Hall-Littlewood polynomials are given by
\begin{equation}
\Psi^{(n_1,...,n_N)}_{U(N)}(x_1,...,x_N;t)= \sum_{\sigma \in S_{N}} x_{\sigma(1)}^{n_1}...x_{\sigma(N)}^{n_N}\prod_{1 \leq i < j \leq N}\frac{1-t^2x_{\sigma(i)}^{-1}x_{\sigma(j)}}{1-x_{ \sigma(i)}^{-1}x_{\sigma(j)}},
\end{equation}
while the factor $\delta(\textbf{n})$ reads
\begin{equation}
\delta(\textbf{n}) = \sum_{j=1}^{N}(N+1-2j)n_j.
\end{equation}
The function $K_{\rho}(\textbf{x};t)$ depends on the particular partition $\rho$ that have been considered. Explicitly it reads
\begin{equation}
K_{\rho}(\textbf{x};t) = \prod_{i=1}^{\textrm{length}(\rho^T)}\prod_{j,k=1}^{\rho_i^T}\frac{1}{1-a_j^i\bar{a}_k^i},
\end{equation}
where $\rho^T$ is the transpose of the partition $\rho$ and the two factors $a_j^i$ and $\bar{a}_k^i$ are given by
\begin{equation}\begin{split}
& a_j^{i} = x_jt^{\rho_j-i+1}, \ \ \ \  \ i=1,...,\rho_j, \\
& \bar{a}_k^{i} = x_k^{-1}t^{\rho_k-i+1}, \ \  \ i=1,,...,\rho_k,
\end{split}\end{equation}
these factors are associated to each box in the Young tableau. Finally $\textbf{w}_{r}$ denotes the weights of the $SU(2)$ representation of dimension $r$
\begin{equation}
\textbf{w}_r = (r-1,r-3,...,3-r,1-r),
\end{equation}
therefore the notation $t^{\textbf{w}_r}$ stands for the vector
\begin{equation}
t^{\textbf{w}_r} = (t^{r-1},t^{r-3},...,t^{3-r},t^{1-r}).
\end{equation}
The formula (\ref{eq:hsp}) admits a generalization for other classical gauge groups. We refer the interested reader to \cite{Cremonesi:2014kwa} for a discussion of these cases.

\subsection{Mirrors of 3d Sicilian theories and the gluing formula}
\label{subsec:glue}
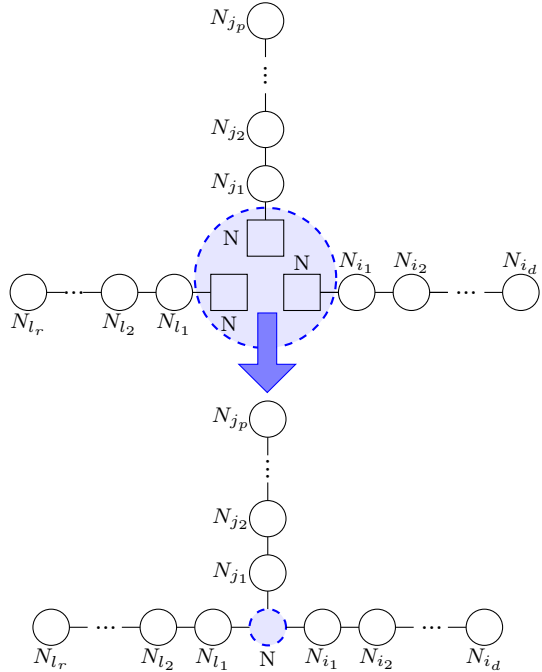
\begin{figure}[h!]
\center{
\begin{tikzpicture}[scale=0.60]


\draw[blue,thick,dashed,fill=blue!10] (-0.8,0.32) circle (1.55cm);
 
 
\draw (0.4,0) -- (0.8,0);
\draw (-0.4,-0.4) rectangle (0.4,0.4);
\draw (0,0.65) node {\footnotesize{N}};

\draw (1.6,0) -- (2,0);
\draw (1.2,0) circle (0.4cm);
\draw (1.2,0.65) node {\footnotesize{$N_{i_1}$}};

\draw (2.8,0) -- (3.2,0);
\draw (2.4,0) circle (0.4cm);
\draw (2.4,0.65) node {\footnotesize{$N_{i_2}$}};

\draw (2.8,0) -- (3.2,0);
\draw (3.6,0) node {...};

\draw (4,0) -- (4.4,0);
\draw (4.8,0) circle (0.4cm);
\draw (4.8,0.65) node {\footnotesize{$N_{i_d}$}};



\draw (0.4,0) -- (0.8,0);
\draw (-2,-0.4) rectangle (-1.2,0.4);
\draw (-1.6,-0.7) node {\footnotesize{N}};

\draw (-2,0) -- (-2.4,0);
\draw (-2.8,0) circle (0.4cm);
\draw (-2.8,-0.7) node {\footnotesize{$N_{l_1}$}};

\draw (-3.2,0) -- (-3.6,0);
\draw (-4,0) circle (0.4cm);
\draw (-4,-0.7) node {\footnotesize{$N_{l_2}$}};

\draw (-4.4,0) -- (-4.8,0);
\draw (-5,0) node {...};

\draw (-5.2,0) -- (-5.6,0);
\draw (-6.0,0) circle (0.4cm);
\draw (-6,-0.7) node {\footnotesize{$N_{l_r}$}};



\draw (-0.8,1.6) -- (-0.8,2);
\draw (-1.2,0.8) rectangle (-0.4,1.6);
\draw (-1.55,1.2) node {\footnotesize{N}};

\draw (-0.8,2.8) -- (-0.8,3.2);
\draw (-0.8,2.4) circle (0.4cm);
\draw (-1.6,2.4) node {\footnotesize{$N_{j_1}$}};

\draw (-0.8,4) -- (-0.8,4.4);
\draw (-0.8,3.6) circle (0.4cm);
\draw (-1.6,3.6) node {\footnotesize{$N_{j_2}$}};

\draw (-0.8,5.2) -- (-0.8,5.6);
\draw (-0.8,4.8) node[rotate=90] {...};

\draw (-0.8,6) circle (0.4cm);
\draw (-1.6,6) node {\footnotesize{$N_{j_p}$}};



\node[single arrow, draw, align=center, xshift=-1.3em, yshift=-2.1em, rotate=-90, minimum height=3.0em, blue,fill=blue!50](potok1){};


\draw (-0.75,-3.2) -- (-0.75,-3.6);
\draw (-0.75,-2.8) circle (0.4cm);
\draw (-1.55,-2.8) node {\footnotesize{$N_{j_p}$}};

\draw (-0.75,-4.2) -- (-0.75,-4.6);
\draw (-0.75,-3.9) node[rotate=90] {...};

\draw (-0.75,-5.0) circle (0.4cm);
\draw (-1.55,-5.0) node {\footnotesize{$N_{j_2}$}};
\draw (-0.75,-5.4) -- (-0.75,-5.8);

\draw (-0.75,-6.2) circle (0.4cm);
\draw (-1.55,-6.2) node {\footnotesize{$N_{j_1}$}};
\draw (-0.75,-6.6) -- (-0.75,-7);

\draw[blue,thick,dashed,fill=blue!10] (-0.75,-7.4) circle (0.4cm);
\draw (-0.75,-8.1) node {\footnotesize{N}};


\draw (-1.15,-7.4) -- (-1.55,-7.4);
\draw (-1.95,-7.4) circle (0.4cm);
\draw (-1.95,-8.1) node {\footnotesize{$N_{l_1}$}};

\draw (-2.35,-7.4) -- (-2.75,-7.4);
\draw (-3.15,-7.4) circle (0.4cm);
\draw (-3.15,-8.1) node {\footnotesize{$N_{l_2}$}};

\draw (-3.55,-7.4) -- (-3.95,-7.4);
\draw (-4.3,-7.4) node {...};

\draw (-4.7,-7.4) -- (-5.1,-7.4);
\draw (-5.5,-7.4) circle (0.4cm);
\draw (-5.5,-8.1) node {\footnotesize{$N_{l_r}$}};


\draw (-0.35,-7.4) -- (0.05,-7.4);
\draw (0.45,-7.4) circle (0.4cm);
\draw (0.45,-8.1) node {\footnotesize{$N_{i_1}$}};

\draw (0.85,-7.4) -- (1.25,-7.4);
\draw (1.65,-7.4) circle (0.4cm);
\draw (1.65,-8.1) node {\footnotesize{$N_{i_2}$}};

\draw (2.05,-7.4) -- (2.45,-7.4);
\draw (2.85,-7.4) node {...};

\draw (3.20,-7.4) -- (3.60,-7.4);
\draw (4,-7.4) circle (0.4cm);
\draw (4,-8.1) node {\footnotesize{$N_{i_d}$}};

\end{tikzpicture}
}
\caption{\label{fig:glue}Graphical representation of the \textit{gluing technique}, in the particular case in which only three $T_{\rho_i}[SU(N)]$ theories are involved.}
\end{figure}

In this section we review the formula for the Hilbert Series of Coulomb branch of the mirrors of $3d$ Sicilian theories \cite{Cremonesi:2014uva}. These theories are be obtained starting from the $6d$ (2,0) theory with symmetry group $G$ performing a compactification over a punctured Riemann surface times a circle. Therefore these theories can be understood as the mirrors of the theories on M5-branes wrapping a circle times a punctured sphere $\rho_1,...,\rho_n$ \cite{Benini:2010uu,Nishioka:2011dq}. These theories are described by a \textit{star-shaped} quiver gauge theory. This is a quiver diagram with $n$-arms all connected trough a central node. An example of star-shaped quiver with three arms is reported in figure \ref{fig:glue}. Each arm $i$ of the quiver diagram is associated to a different $T_{\rho_i}(G)$ theory. 

A formula that allows to obtain the Coulomb branch Hilbert Series for these class of theories have been introduced in \cite{Cremonesi:2014uva}. It was demonstrated that the Hilbert Series for the Coulomb branch of the mirrors of the $3d$ Sicilian theories is obtained by ``gluing" together the Hilbert Series for the different $T_{\rho}(G)$ theories associated to the different arms of the quiver diagram and that share the same global symmetry group. Specifically given a set of theories $\{T_{\rho_1}(G),...,T_{\rho_e}(G) \}$ all with the same global symmetry group $G$ we can construct the corresponding mirror theory gauging the common centerless flavor symmetry $G/Z(G)$. Henceforth we restrict our attention to the particular case in which $G=SU(N)$ and we refer the reader to \cite{Cremonesi:2014uva} for the discussion of the general case. In this particular case the Hilbert Series of the resulting theory reads
\begin{equation}
\label{hsglue}
\begin{split}
& \textrm{HS}(t;\textbf{x}^{(1)},...,\textbf{x}^{(e)}) = \\
& = \sum_{n_1 \geq ... n_N \geq 0} \prod_{j=1}^{e} \textrm{HS}[T_{\rho_j}(SU(N))](t;\textbf{x}^{(j)};n_1,...,n_N)t^{\delta(\textbf{n})}(1-t^2)P_{U(N)}(t;n_1,...n_N),
\end{split}
\end{equation}
we observe that the Hilbert Series of the resulting theory is obtained multiplying the Hilbert Series of the building blocks of the star-shaped quiver (i.e. the different $T_{\rho_i}(G)$ theories) and then summing over the monopoles of the gauged $SU(N)$ group. See figure \ref{fig:glue} for a graphical representation of this formula.
In this paper we consider only star-shaped quiver with three arms. Remarkably the global symmetry group $G_{global}$ of these theories can be extracted directly from their quiver diagram using the following procedure \cite{Gaiotto:2008ak,Chacaltana:2010ks}:
\begin{enumerate}
\item Identify all the so called \textit{balanced-nodes} of the quiver (i.e. a node for which the sum over the ranks $k_i$ of the adjacent nodes is equal to $2k$, where $k$ is  the rank of the node taken in consideration).
\item If all nodes are gauged ungauge a $U(1)$ by choice.  
\item Then the balanced nodes will form the Dynkin diagram of the semi-simple part of $G_{global}$. The abelian part of the global symmetry group is $U(1)^{k-1}$, where $k$ is the number of unbalanced nodes of the quiver.
\end{enumerate}
In the following we apply the above prescription to the five families of quiver gauge theories reported in section \ref{sec:results}.

\section{Coulomb branch and nilpotent orbits}
\label{sec:nilpotent}
Following \cite{Cabrera:2016vvv} in this section we summarize the basic information regarding nilpotent orbits that will be relevant in the following parts of this paper. Recently it has been understood that the Coulomb branch and the Higgs branch of a $3d$ $\mathcal{N}=4$ theory can be related to nilpotent orbits \cite{Cabrera:2016vvv}. As a matter of fact Namikawa's theorem \cite{2016arXiv160306105N} states that if the Coulomb branch or the Higgs branch is finitely generated by operators with spin one under the $SU(2)_R$ symmetry group then this space is the closure of a nilpotent orbit of the isometry group of the algebra. Spaces with generators with spin higher than one can be thought as extensions of  closures of nilpotent orbits. Moreover the generators with spin 1 transform in the adjoint representation of an isometry group of the variety \cite{1992math......4227B}.

Let's focus on nilpotent orbits of the Lie algebra $\textbf{g}=\textbf{sl}_n$. These are in a one to one correspondence with the partitions of $n$, this is a n-tuple $\lambda=(\lambda_1,\lambda_2...\lambda_n)$ that satisfies 
\begin{equation}
\lambda_1 \geq \lambda_2 \geq ... \geq \lambda_n, \ \ \ \textrm{and} \ \ \ \sum_{i=1}^{k} \lambda_i = n.
\end{equation}
An \textit{elementary Jordan block} of order m is a $m \times m$ matrix,
\begin{equation}
\label{eq:jb}
J^{m} = \left( \begin{array}{cccccc}
0 & 1 & 0 & ... & 0 & 0 \\ 
0 & 0 & 1 & ... & 0 & 0 \\ 
\vdots & \vdots  & \vdots & \ddots & \vdots & \vdots \\ 
0 & 0 & 0 & ... & 0 & 1 \\ 
0 & 0 & 0 & ... & 0 & 0
\end{array}   \right),
\end{equation}  
given a partition $\lambda = (\lambda_1,...,\lambda_n)$ we can introduce the matrix $X^{\lambda}$ associated to the partition $\lambda$
\begin{equation}
X^{\lambda} = \oplus_{i} J^{\lambda_i}.
\end{equation}
The \textit{nilpotent orbit} $\mathcal{O}_{\lambda}$ corresponding to the partition $\lambda$ is obtained as
\begin{equation}
\mathcal{O}_{\lambda} = G_{adj} \cdot X^{\lambda},
\end{equation}
where $G_{adj}$ denotes the action of the adjoint group. We refer the reader to \cite{nilpotent} for more details regarding nilpotent orbits.

In the following sections, in order to outline the relation between closures of nilpotent orbits and Coulomb branch, we consider the PLog expansion of the Hilbert Series and we set equal to zero all the generators with spin higher than one. For each theory we find nilpotent generators (with spin equal to one) that can be represented by a $N \times N$ nilpotent matrix  $M$, such that $M^2=0$, and satisfying the \textit{Jordan condition}
\begin{equation}
\label{eq:jordan}
\textrm{Tr}[M^p] = 0 \ \forall \ p \in \mathbb{N} \ \Leftrightarrow \ \textrm{all eigenvalues of} \ M \ \textrm{are} \ 0.
\end{equation}
This implies that the nilpotent matrix $M$ can be only decomposed in the elementary Jordan blocks $J^1$ and $J^2$. Therefore the rank of the matrix $M$ must satisfy the constraint
\begin{equation}
\textrm{rank}[M] \leq \frac{N}{2}.
\end{equation}
The above information allow to relate $M$ to the closure of a nilpotent orbit of $\textbf{sl}_n$.   This also implies that the nilpotent orbit which results by setting the extra generators to zero is at most of type $(2^k,1^{N-2k})$ \footnote{We use the shorthand notation $(p^{k})=(\underbrace{p,...,p}_{k \ \ times})$.} with $k < N/2$ and $k \in \mathbb{N}$.

\section{Overview of the results}
\label{sec:results}
In this section we report our main results, i.e. the general expression of the HWG for the five families of quiver gauge theories that have been considered. We focused our attention on three families of quiver gauge theories with unitary global symmetry group (reported in section \ref{subsec:unitary}) and two families with orthogonal global symmetry group (reported in section \ref{subsec:orth}). We report the quiver diagram and the corresponding global symmetry group of each theory.\footnote{Henceforth the balanced nodes of all the quiver diagrams are marked in red.}
\subsection{Theories with unitary global symmetry}
\label{subsec:unitary}
\subsubsection{Theories with $G_{global} = SU(2)\times SU(2N)$}
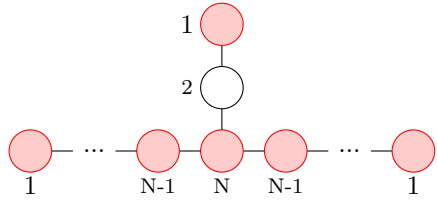
\begin{figure}[h]
\center{
\begin{tikzpicture}[scale=0.70]

\draw (0.4,0) -- (0.8,0);
\draw[red, fill=red!20] (0,0) circle (0.4cm);
\draw (0,-0.65) node {\footnotesize{N}};

\draw (1.6,0) -- (2,0);
\draw[red, fill=red!20] (1.2,0) circle (0.4cm);
\draw (1.2,-0.65) node {\footnotesize{N-1}};

\draw (2.8,0) -- (3.2,0);
\draw (2.4,0) node {...};

\draw[red, fill=red!20] (3.6,0) circle (0.4cm);
\draw (3.6,-0.65) node {1};

\draw (-0.4,0) -- (-0.8,0);
\draw[red,fill=red!20] (-1.2,0) circle (0.4cm);
\draw (-1.2,-0.65) node {\footnotesize{N-1}};

\draw (-1.6,0) -- (-2,0);
\draw (-2.4,0) node{...};

\draw (-2.8,0) -- (-3.2,0);
\draw[red,fill=red!20] (-3.6,0) circle(0.4cm);
\draw (-3.6,-0.65) node {1};

\draw (0,0.4) -- (0,0.8);
\draw (0,1.2) circle (0.4cm);
\draw (-0.65,1.2) node {\footnotesize{2}};

\draw (0,1.6) -- (0,2.0);
\draw[red, fill=red!20] (0,2.4) circle (0.4cm);
\draw (-0.65,2.4) node {1};

\end{tikzpicture}
}
\caption{\label{fig:su2su2n} Quiver diagram with $SU(2N) \times SU(2)$ global symmetry group.}
\end{figure}

These theories are described by the quiver diagram reported in figure \ref{fig:su2su2n}

Note that the case $N=3$ is special since the global symmetry is enhanced to $E_6$. The HWG for this case was already discussed in  \cite{Hanany:2015hxa} (see the first line of table 10 of \cite{Hanany:2015hxa}).
In general the HWG for this class of theories reads
\begin{equation}
\label{eq:hwgsu2nsu2}
\resizebox{.94\hsize}{!}{$
\textrm{HWG}_{SU(2) \times SU(2N)}(t;\nu,\mu_i)= 
\textrm{PE}\left[\nu^2t^2 +t^4 +\nu\mu_Nt^{N-1} + \nu\mu_Nt^{N+1} +  \sum_{i=1}^{N}\mu_i\mu_{2N-i}t^{2i}  -\nu^2\mu_{N}^2t^{2N+2} \right],  
$}
\end{equation}
where $\nu$ is the highest weight of the $SU(2)$ subgroup and the various $\mu_j$ denote the highest weights of the $SU(2N)$ subgroup. 

\subsubsection{Theories with $G_{global} = SU(2N)$}

\begin{figure}[h!]
\center{
\begin{tikzpicture}[scale=0.70]

\draw (0.4,0) -- (0.8,0);
\draw[red, fill=red!20] (0,0) circle (0.4cm);
\draw (0,-0.65) node {\footnotesize{N}};

\draw (1.6,0) -- (2,0);
\draw[red, fill=red!20] (1.2,0) circle (0.4cm);
\draw (1.2,-0.65) node {\footnotesize{N-1}};

\draw (2.8,0) -- (3.2,0);
\draw (2.4,0) node {...};

\draw[red, fill=red!20] (3.6,0) circle (0.4cm);
\draw (3.6,-0.65) node {1};

\draw (-0.4,0) -- (-0.8,0);
\draw[red,fill=red!20] (-1.2,0) circle (0.4cm);
\draw (-1.2,-0.65) node {\footnotesize{N-1}};

\draw (-1.6,0) -- (-2,0);
\draw (-2.4,0) node{...};

\draw (-2.8,0) -- (-3.2,0);
\draw[red,fill=red!20] (-3.6,0) circle(0.4cm);
\draw (-3.6,-0.65) node {1};

\draw (0,0.4) -- (0,0.8);
\draw (0,1.2) circle (0.4cm);
\draw (-0.65,1.2) node {\footnotesize{2}};


\end{tikzpicture}
}
\caption{\label{fig:su2n}Quiver diagram with $SU(2N)$ global symmetry group.}
\end{figure}
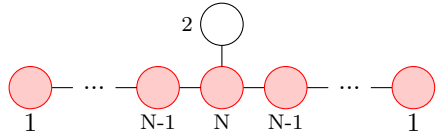

These theories are described by the quiver diagram reported in figure \ref{fig:su2n}.

Note that the case $N=4$ is special since the global symmetry is enhanced to $E_7$. The HWG for this case was already discussed in  \cite{Hanany:2015hxa} (see the fourth line of table 10 of \cite{Hanany:2015hxa}). In general the HWG for this class of theories reads
\begin{equation}
\label{eq:hwgsu2n}
\textrm{HWG}_{SU(2N)}(t;\mu_i) = \textrm{PE}\left[t^4 + \sum_{i=1}^{N-1}\mu_i\mu_{2N-i}t^{2i} + \mu_Nt^{N-2} + \mu_Nt^{N} \right],
\end{equation}
where $\mu_j$ are the highest weights of the $SU(2N)$ group.

\subsubsection{The mirror of the $(k)-[2N]$ theory, $G_{global} = SU(2N)$}

\begin{figure}[h!]
\center{
\begin{tikzpicture}[scale=0.70]


\draw (4.0,0) -- (4.4,0);
\draw (4.8,0) node{...};
\draw (5.2,0) -- (5.6,0);

\draw[red, fill=red!20] (6,0) circle (0.4cm);
\draw (6,-0.65) node {2};
\draw (6.4,0) -- (6.8,0);

\draw[red, fill=red!20] (7.2,0) circle (0.4cm);
\draw (7.2,-0.65) node {1};


\draw (0.4,0) -- (0.8,0);
\draw[red, fill=red!20] (0,0) circle (0.4cm);
\draw (0,-0.65) node {\footnotesize{k}};

\draw (1.2,0) node {...};
\draw (1.6,0) -- (2,0);
\draw[red, fill=red!20] (2.4,0) circle (0.4cm);
\draw (2.4,-0.65) node {\footnotesize{k}};

\draw (2.8,0) -- (3.2,0);
\draw[red, fill=red!20] (3.6,0) circle (0.4cm);
\draw (3.6,-0.65) node {\footnotesize{k}};

\draw (3.6,0.4) -- (3.6,0.8);
\draw (3.25,0.8) rectangle (3.95,1.5);
\draw (4.15,1.2) node {\footnotesize{1}};

\draw (-1.2,0) -- (-1.2,0.8);
\draw (-1.55,0.8) rectangle (-0.85,1.5);
\draw (-1.75,1.2) node {\footnotesize{1}};

\draw (-0.4,0) -- (-0.8,0);
\draw[red,fill=red!20] (-1.2,0) circle (0.4cm);
\draw (-1.2,-0.65) node {\footnotesize{k}};

\draw (-1.6,0) -- (-2,0);
\draw (-2.4,0) node{...};

\draw (-2.8,0) -- (-3.2,0);
\draw[red,fill=red!20] (-3.6,0) circle(0.4cm);
\draw (-3.6,-0.65) node {2};

\draw (-4,0) -- (-4.4,0);
\draw[red,fill=red!20] (-4.8,0) circle(0.4cm);
\draw (-4.8,-0.65) node {1};


\draw[decorate,decoration={brace,mirror,amplitude=5pt},xshift=0pt,yshift=0pt]
(-0.2,-0.9) -- (2.8,-0.9) node [black,midway,xshift=0cm,yshift=-0.4cm] 
{\footnotesize $2N-2k-1 \ \textrm{nodes}$};

\end{tikzpicture}
}
\caption{\label{fig:mirror}Quiver diagram of the mirror of the  $(k)-[2N]$ theory, with $k<N$.}
\end{figure}
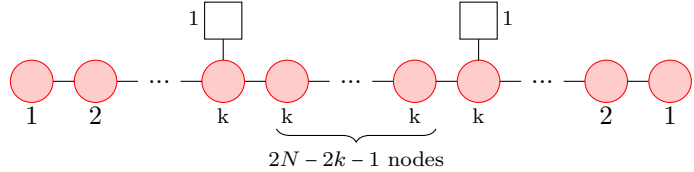

The mirror of the (k)-[2N] theory is described by the quiver reported in figure \ref{fig:mirror}

The corresponding HWG reads
\begin{equation}
\label{eq:hwgk2n}
\textrm{HWG}_{N,k}(t;\mu_i) = \textrm{PE}\left[\sum_{i=1}^{k} \mu_i\mu_{2N-i}t^{2i}\right],
\end{equation}
where $\mu_j$ are the highest weights of the $SU(2N)$ group.
The expression (\ref{eq:hwgk2n}) for the HWG was already found in \cite{Ferlito:2016grh}. Moreover the relation between the Coulomb branch of this class of theories and closure of nilpotent orbits was analysed \cite{Cabrera:2016vvv}.

We test the expressions of the HWG (\ref{eq:hwgsu2nsu2}), (\ref{eq:hwgsu2n}) and (\ref{eq:hwgk2n}) in section \ref{sec:hscomp} considering different values of $N$.

\subsection{Theory with orthogonal global symmetry}
\label{subsec:orth}
\subsubsection{Theories with $G_{Global} = SO(4N+6)\times U(1)$}
\begin{figure}[h!]
\center{
\begin{tikzpicture}[scale=0.70]

\draw (0.4,0) -- (0.8,0);
\draw[red, fill=red!20] (0,0) circle (0.4cm);
\draw (0,-0.65) node {\footnotesize{2N+1}};

\draw (1.6,0) -- (2,0);
\draw[red, fill=red!20] (1.2,0) circle (0.4cm);
\draw (1.2,-0.65) node {\footnotesize{2N}};

\draw (2.8,0) -- (3.2,0);
\draw[red, fill=red!20] (2.4,0) circle (0.4cm);
\draw (2.4,-0.65) node {\footnotesize{2N-1}};

\draw (3.5,0) node {...};

\draw(3.8,0) -- (4.2,0);
\draw[red, fill=red!20] (4.6,0) circle (0.4cm);
\draw (4.6,-0.65) node {1};

\draw (-0.4,0) -- (-0.8,0);
\draw[red,fill=red!20] (-1.2,0) circle (0.4cm);
\draw (-1.2,-0.65) node {\footnotesize{N+1}};

\draw (-1.6,0) -- (-2,0);
\draw (-2.4,0) circle (0.4cm);
\draw (-2.4,-0.65) node {1};

\draw (0,0.4) -- (0,0.8);
\draw[red,fill=red!20] (0,1.2) circle (0.4cm);
\draw (-0.85,1.2) node {\footnotesize{N+1}};

\draw (0,1.6) -- (0,2.0);
\draw (0,2.4) circle (0.4cm);
\draw (-0.65,2.4) node {1};

\end{tikzpicture}
}
\caption{\label{fig:orth1} Quiver diagram with $SO(4N+6) \times U(1)$ global symmetry group. Note that when $N=1$ the global symmetry is enhanced to $E_6$.}
\end{figure}
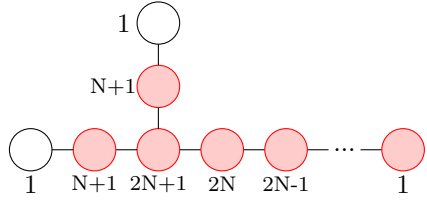

These theories are described by the quiver diagram reported in figure \ref{fig:orth1} 

We conjecture the following HWG for this class of theories
\begin{equation}
\label{eq:hwgso4n6u1}
\textrm{HWG}_{SO(4N+6)\times U(1)}(t;\mu_i,q)=\textrm{PE}\left[t^2 + \sum_{i=1}^{N}\mu_{2i}t^{2i} + \frac{\mu_{2N+3}t^{N+1}}{q} + q\mu_{2N+2}t^{N+1} \right],
\end{equation}
where $q$ is the highest weight of the $U(1)$ subgroup while the various $\mu_i$ are the highest weights of the $SO(4N+6)$ group.

\subsubsection{Theories with $G_{global} = SO(4N+4)$, with $N \geq 3$}
\begin{figure}[h!]
\center{
\begin{tikzpicture}[scale=0.70]

\draw (0.4,0) -- (0.8,0);
\draw[red, fill=red!20] (0,0) circle (0.4cm);
\draw (0,-0.85) node {\footnotesize{2N}};

\draw (1.6,0) -- (2,0);
\draw[red, fill=red!20] (1.2,0) circle (0.4cm);
\draw (1.2,-0.85) node {\footnotesize{2N-1}};

\draw (2.8,0) -- (3.2,0);
\draw (2.4,0) node {...};

\draw[red, fill=red!20] (3.6,0) circle (0.4cm);
\draw (3.6,-0.85) node {1};

\draw (-0.4,0) -- (-0.8,0);
\draw[red,fill=red!20] (-1.2,0) circle (0.4cm);
\draw (-1.2,-0.85) node {\footnotesize{N+1}};

\draw (-1.6,0) -- (-2,0);
\draw (-2.4,0) circle (0.4cm);
\draw (-2.4,-0.85) node {2};

\draw (0,0.4) -- (0,0.8);
\draw[red,fill=red!20] (0,1.2) circle (0.4cm);
\draw (-0.85,1.2) node {\footnotesize{N}};

\end{tikzpicture}
}
\caption{quiver diagram with $SO(4N+4)$ global symmetry group.\label{fig:so4n}}
\end{figure}
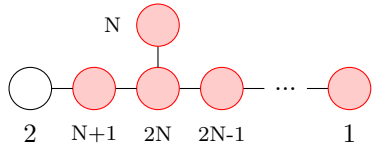

These theories are described by the quiver diagram reported in figure \ref{fig:so4n} \footnote{We must require $N \geq 3$ in order to ensure that all the partitions are non-increasing.}

Note that the case $N=3$ is special since the global symmetry is enhanced to $E_8$. This case was already discussed in \cite{Hanany:2015hxa} (see the sixth line of table 10 of \cite{Hanany:2015hxa}). We conjecture the following HWG for this class of theories
\begin{equation}
\label{eq:hwgso4n4}
\textrm{HWG}_{SO(4N+4)}(t;\mu_i) = \textrm{PE}\left[ \sum_{i=1}^{N}\mu_{2i}t^{2i} + t^4 + \mu_{2N+2}(t^{N-1}+t^{N+1})  \right] \ ,
\end{equation}
where $\mu_i$ are the highest weights of $SO(4N+4)$.

We test the expression (\ref{eq:hwgso4n6u1}) and the expression (\ref{eq:hwgso4n4}) in section \ref{sec:globalort}.

\section{Theories with unitary global symmetry group} 
\label{sec:hscomp}
In this section we test the expressions of the HWG (\ref{eq:hwgsu2nsu2}) and (\ref{eq:hwgsu2n}) for several values of the integer $N$ characterizing the quiver gauge theory. For each theory we report the expression of the corresponding HWG, the first orders of the expansion of the HS and of the Pletystic logarithm, the first generators and their relations. Finally we analyse the connection between the Coulomb branch and  closure of nilpotent orbits. We refer the reader to appendix \ref{app:quiver} for further details regarding these computations.
\subsection{$N=3$, \ $G_{global}$ = $E_6 \supset SU(2) \times SU(6) $}
\label{subsec:e6}
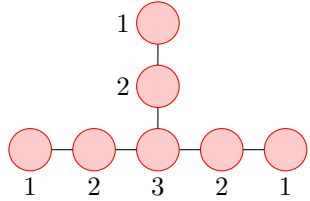
\begin{figure}[h]
\center{
\begin{tikzpicture}[scale=0.70]

\draw (0.4,0) -- (0.8,0);
\draw[red, fill=red!20] (0,0) circle (0.4cm);
\draw (0,-0.7) node {3};

\draw (1.6,0) -- (2,0);
\draw[red, fill=red!20] (1.2,0) circle (0.4cm);
\draw (1.2,-0.7) node {2};

\draw[red, fill=red!20] (2.4,0) circle (0.4cm);
\draw (2.4,-0.7) node {1};

\draw (-0.4,0) -- (-0.8,0);
\draw[red, fill=red!20] (-1.2,0) circle (0.4cm);
\draw (-1.2,-0.7) node {2};

\draw (-1.6,0) -- (-2,0);
\draw[red, fill=red!20] (-2.4,0) circle (0.4cm);
\draw (-2.4,-0.7) node {1};

\draw (0,0.4) -- (0,0.8);
\draw[red, fill=red!20] (0,1.2) circle (0.4cm);
\draw (-0.65,1.2) node {2};

\draw (0,1.6) -- (0,2);
\draw[red, fill=red!20] (0,2.4) circle (0.4cm);
\draw (-0.65,2.4) node {1};

\end{tikzpicture}
}
\caption{Quiver diagram with $E_{6}$ global symmetry group. \label{fig:e6}}
\end{figure}

The quiver gauge theory with $E_6$ global symmetry group is reported in figure \ref{fig:e6}.
We perform the computation of corresponding unrefined HS using the formula (\ref{hsglue}). At the lowest orders in the $t$ expansion we find \footnote{Note that this is the Hilbert Series of the reduced moduli space of one  $E_{6}$ instanton on $\mathbb{C}^{2}$, whose computation was also performed in a different way in  \cite{Benvenuti:2010pq}.}
\begin{equation}
\textrm{HS}_{E_6}(t,1,...,1)=1 + 78t^2 +2430t^4 +43758t^6 +537966t^8 +o(t^9).
\end{equation}
At every order in the $t$-expansion we decompose the $E_6$ representations under representation of the global symmetry subgroup $SU(2) \times SU(6)$. This way we find the HWG \footnote{Note that this result is an agreement with the HWG previously found in \cite{Hanany:2015hxa,Benvenuti:2010pq}.}
\begin{equation}\begin{split}
\label{eq:hwge6}
& \textrm{HWG}_{SU(2) \times SU(6)}(t;\nu,\mu_{i})  =   \textrm{PE}[(\nu^2+ \mu_1\mu_5 + \nu\mu_3)t^2 +(1+ \mu_2\mu_4+\mu_3\nu)t^4 + \mu_3^2t^6 -\nu^2\mu_3^2t^8].
\end{split}\end{equation}
This result for the HWG follows the pattern outlined in equation (\ref{eq:hwgsu2nsu2}). Therefore using the above HWG the corresponding HS can be written in terms of $SU(2)$ and $SU(6)$ representations as \footnote{Henceforth the notation $[a;b_1,b_2,...,b_{2N-1}]$ denotes the product between the $[a]$ $SU(2)$ representation and the $[b_1,b_2,...,b_{2N-1}]$  $SU(2N)$ representation.}
\begin{equation}
\label{eq:hssu2su6}
\begin{split}
& \textrm{HS}_{SU(2)\times SU(6)}(t;x,y_i) = 1+ ([0;1,0,0,0,1] + [1;0,0,1,0,0] + [2;0,0,0,0,0])t^2 + ([0;0,0,0,0,0]\\
& + [0;0,1,0,1,0]+ [0;2,0,0,0,2] + [1;0,0,1,0,0]+[1;1,0,1,0,1]+[2;1,0,0,0,1] + [2;0,0,2,0,0]\\
& [3;0,0,1,0,0] + [4;0,0,0,0,0])t^4 + o(t^4). 
\end{split}\end{equation}
The Plethystic logarithm of the Hilbert series (\ref{eq:hssu2su6}) reads
\begin{equation}\begin{split}
\label{eq:plsu2su6}
& \textrm{PLog}[\textrm{HS}_{SU(2) \times SU(6)}(t;x,y_i)] = ([0;1,0,0,0,1]+ [1;0,0,1,0,0] + [2;0,0,0,0,0])t^2 + \\
& - (2[0;0,0,0,0,0] + [0;1,0,0,0,1] + [0;0,1,0,1,0] + [1;1,1,0,0,0] + [1;0,0,0,1,1] \\
& + [1;0,0,1,0,0] + [2;1,0,0,0,1])t^4 + o(t^4).
\end{split}\end{equation}
\subsubsection{The generators and their relations}
At the order $t^2$ of the expansion (\ref{eq:plsu2su6}) we have three generators \footnote{A similar analysis has also been carried out in \cite{Gaiotto:2008nz}.}
\begin{align}
[0;1,0,0,0,1]:& \ \ \ \ \ M^{i_1}_{ \ \ i_2} \ \ \textrm{with} \ \ \textrm{Tr}[M]=0, \\
[1;0,0,1,0,0]:& \ \ \ \ \ N^{[i_1i_2i_3]}_\alpha, \\ 
[2;0,0,0,0,0]:& \ \ \ \ \ C_{\alpha\beta} \ \ \textrm{with} \ \ \textrm{Tr}[C]=0, 
\end{align}
where $i_1,...,i_6=1,... ,6$ are $SU(6)$ indices while $\alpha,\beta=1,2$ are $SU(2)$ indices. The generator $M$ transforms under the adjoint representation of $SU(6)$, the generator $C$ transforms under the adjoint representation of $SU(2)$, while the generator $N^{[i_1i_2i_3]}_{\alpha}$ transforms under the completely antisymmetric representation of $SU(6)$ and under the fundamental representation of $SU(2)$.

At the order $t^4$ there are the following relations \footnote{Where for a tensor with the structure $P_{i[jk]}$ we introduced the curly brackets $\{ \}$ 
\begin{equation}
P_{\{i[jk]\}} = P_{i[jk]} - P_{[i[jk]]}.
\end{equation}}

\begin{align}
[2;1,0,0,0,1]:& \ \ \ \ \ M^{i_1}_{ \ \ j_1}C_{\alpha \beta} + \frac{1}{4}\left(N^{[i_1i_2i_3]}_{(\alpha}N_{[j_1i_2i_3]\beta)}\right) = 0,\label{eq:relort1}\\
[1;1,1,0,0,0]:& \ \ \ \ \ M^{j_1}_{\  \{i_1}N_{j_2j_3 \}j_1 \alpha} = 0,\label{eq:relort2}\\
[1;0,0,0,1,1]:& \ \ \ \ \ M^{\{i_1}_{\ \ j_1}N^{j_2j_3\}j_1}_{\alpha} = 0,\label{eq:relort3}\\ 
[1;0,0,1,0,0]:& \ \ \ \ \ N^{i_1i_2i_3}_{\alpha}C_{\beta \gamma}\epsilon^{\alpha \beta} + M^{[i_1}_{ \ \ j_1}N^{i_2i_3]j_1}_\gamma = 0,\label{eq:relort4}\\ 
[0;0,1,0,1,0]:& \ \ \ \ \ (N^{i_1i_2q}_\alpha N_{j_1j_2q \beta}\epsilon^{\alpha \beta} -4M^{[i_1}_{\ \ [j_1}M^{i_1]}_{\ \ j_2]})\mid_{[0,1,0,1,0]} = 0,\label{eq:rele6rankm}\\ 
[0;1,0,0,0,1]:& \ \ \ \ \ M^{i_1}_{\ \ j_1}M^{j_1}_{\ \ i_2} -\frac{1}{6}\delta^{i_1}_{\ \ i_2}M^{j_2}_{\ \ j_3}M^{j_3}_{\ \ j_2}  = 0,\label{eq:rele6m}\\ 
[0;0,0,0,0,0]:& \ \ \ \ \ N^{i_1i_2i_3}_{\alpha}N_{i_1i_2i_3 \ \beta}\epsilon^{\alpha\beta} + 24C_{\alpha\beta}C_{\gamma\delta}\epsilon^{\alpha\beta}\epsilon^{\gamma\delta} = 0,\label{eq:rele6c1}\\ 
[0;0,0,0,0,0]:& \ \ \ \ \ M^{i_1}_{\ \ j_1}M^{j_1}_{ \ \ i_1} + 3C_{\alpha\beta}C_{\gamma\delta}\epsilon^{\alpha\beta}\epsilon^{\gamma\delta} = 0.\label{eq:rele6c2}
\end{align}
There are three generators with spin 1. The relation (\ref{eq:rele6m}) is satisfied if $M$ is nilpotent and if it satisfies the condition (\ref{eq:jordan}). The relations (\ref{eq:rele6c1})-(\ref{eq:rele6c2}) are satisfied if $C$ is nilpotent and if $N^{ijk}_{\alpha}N_{ijk\beta}\epsilon^{\alpha\beta}=0$.
The reduced moduli space of 1-instanton of $E_6$ is identified with the closure of the minimal nilpotent orbit of $E_6$ \cite{kronheimer1990}. The previous analysis suggests that this space can be decomposed in submanifolds. As a matter of fact the nilpotent operator $M$ is related to the closure of the minimal nilpotent orbit of $SU(6)$, i.e. to the reduced moduli space of 1-instanton of $SU(6)$. On the other hand the operator $C$ is related to the closure of the minimal nilpotent orbit of $SU(2)$, i.e. to the reduced moduli space of 1-instanton of $SU(2)$.
\subsection{$N=4$, $G_{global}$ = $E_{7} \supset SU(8)$}
\begin{figure}
\center{
\begin{tikzpicture}[scale=0.7]

\draw (0.4,0) -- (0.8,0);
\draw[red, fill=red!20] (0,0) circle (0.4cm);
\draw (0,-0.7) node {4};

\draw (1.6,0) -- (2,0);
\draw[red, fill=red!20] (1.2,0) circle (0.4cm);
\draw (1.2,-0.7) node {3};

\draw (2.8,0) -- (3.2,0);
\draw[red, fill=red!20] (2.4,0) circle (0.4cm);
\draw (2.4,-0.7) node {2};

\draw[red, fill=red!20] (3.6,0) circle (0.4cm);
\draw (3.6,-0.7) node {1};

\draw (-0.4,0) -- (-0.8,0);
\draw[red, fill=red!20] (-1.2,0) circle (0.4cm);
\draw (-1.2,-0.7) node {3};

\draw (-1.6,0) -- (-2,0);

\draw (-2.8,0) -- (-3.2,0);
\draw[red, fill=red!20] (-2.4,0) circle (0.4cm);
\draw (-2.4,-0.7) node {2};

\draw[red, fill=red!20] (-3.6,0) circle (0.4cm);
\draw (-3.6,-0.7) node {1};

\draw (0,0.4) -- (0,0.8);
\draw[red, fill=red!20] (0,1.2) circle (0.4cm);
\draw (-0.65,1.2) node {2};

\end{tikzpicture}
}
\caption{Quiver diagram with $E_{7}$ global symmetry group. \label{fig:e71}}
\end{figure}
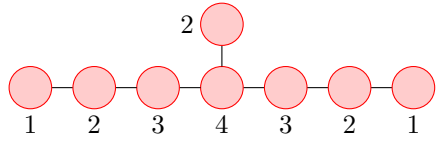

The quiver gauge theory with $E_7$ global symmetry group is reported in figure \ref{fig:e71}.
The lowest orders of the $t$ expansion of the corresponding unrefined HS read \footnote{Note that this is equal to the HS of the reduced moduli space of one-instanton of $E_7$ on $\mathbb{C}^{2}$. This HS has been already computed in \cite{Benvenuti:2010pq}.}
\begin{equation}
\textrm{HS}_{E_7}(t;1,...,1) = 1 +133t^2+ 7371t^{4} + 238602t^{6} +o(t^{8}).
\end{equation}
We decompose the representations of the global symmetry group $E_7$ under representations of its $SU(8)$ subgroup. This way we get the HWG \footnote{This result agrees with the HWG found previously in \cite{Hanany:2015hxa}.}
\begin{equation}
\label{eq:hwgsu8}
\begin{split}
 \textrm{HWG}_{SU(8)}(t;\mu_{i})=  \textrm{PE}[(\mu_1\mu_7+\mu_4)t^2 +(1+\mu_4+\mu_2\mu_6)t^4 + \mu_3\mu_5t^6].
\end{split}
\end{equation}
This result for the HWG follows the pattern outlined in equation (\ref{eq:hwgsu2n}). Therefore using the above HWG the corresponding HS can be written in terms of  $SU(8)$ representations as
\begin{equation}\begin{split}
\label{eq:hssu8}
& \textrm{HS}_{SU(8)}(t;y_i) =  1 + ([1,0,0,0,0,0,1] + [0,0,0,1,0,0,0])t^2 + ([1,0,0,1,0,0,1] + [0,1,0,0,0,1,0] +\\ 
& [0,0,0,1,0,0,0] + [0,0,0,0,0,0,0] + [2,0,0,0,0,0,2] + [0,0,0,2,0,0,0])t^4 + o(t^4),
\end{split}\end{equation}
the Plethystic logarithm of the HS
(\ref{eq:hssu8}) reads
\begin{equation}\begin{split} 
\label{eq:plsu8}
& \textrm{PLog}[\textrm{HS}_{SU(8)}(t;y_i)] =  ([1,0,0,0,0,0,1]+[0,0,0,1,0,0,0])t^2  -([0,0,0,0,0,0,0] + [1,0,0,0,0,0,1] \\
& + [0,1,0,0,0,1,0] + [1,0,1,0,0,0,0] + [0,0,0,0,1,0,1])t^4 +o(t^4).
\end{split}\end{equation}

\subsubsection{The generators and their relations}
At the order $t^2$ of the expansion of the PLog (\ref{eq:plsu8}) there are two generators
\begin{align}
[1,0,0,0,0,0,1]:& \ \ \ M^{i_1}_{ \ \ i_2} \ \ \ \textrm{and} \ \ \ \textrm{Tr}[M]=0, \\
[0,0,0,1,0,0,0]:& \ \ \ N^{[i_1i_2i_3i_4]}, 
\end{align}
where $i_1,...,i_4=1,...,8$ are $SU(8)$ indices. The operator $M^{i_1}_{ \ \ i_2}$ transforms under the adjoint representation of $SU(8)$, while the operator $N^{[i_1i_2i_3i_4]}$ transforms under the representation \textbf{70} of $SU(8)$. At the order $t^4$ there are five relations\footnote{For a generic tensor with the structure $P_{i[jkl]}$ we define the projection to the irrep \textbf{378} as \begin{equation}
P_{\{i[jkl]\}} \equiv P_{i[jkl]} -P_{[ijkl]} .
\end{equation}
}

\begin{align}
[1,0,0,0,0,0,1] + [0,0,0,0,0,0,0]:& \ \ \ M^{i_1}_{ \ \ i_2}M^{i_2}_{ \ \ i_3} + \delta^{i_1}_{\ \ i_3}(N^{j_1j_2j_3j_4}N_{j_1j_2j_3j_4}) = 0  \label{eq:rel1}\\
[0,1,0,0,0,1,0]:& \ \ \ \left(M^{[i_{1}}_{\ \ [j_{1}}M^{i_{2}]}_{\ \  j_{2}]} + c_1N^{i_1i_2i_3i_4}N_{j_1j_2i_3i_4}\right)|_{\textbf{720}} = 0, \label{eq:rel2}\\
[1,0,1,0,0,0,0]:& \ \ \ M_{\ \ j_1}^{i_1}N_{i_1i_2i_3i_4}|_{\overline{\textbf{378}}} = M_{\ \ \{ j_1}^{i_1}N_{i_2i_3i_4\}i_1} = 0, \label{eq:rel31a}\\
[0,0,0,0,1,0,1]:& \ \ \ M^{j_1}_{\ \ i_1}N^{i_1i_2i_3i_4}|_{\textbf{378}} = M^{\{ j_1}_{\ \ \ i_1}N^{i_2i_3i_4\}i_1} = 0, \label{eq:rel31b}
\end{align}
where $c_1 \ \in \Re$. The expression (\ref{eq:hssu8}) is the power series expansion of the first orders of the Hilbert Series of the reduced moduli space of 1-instanton of $E_7$. This space is equal to the  minimal nilpotent orbit of $E_7$ \cite{kronheimer1990}. However it's interesting to analyse the decomposition of $E_7$ representations under $SU(8)$ representations and interpret the corresponding relations in terms of $SU(8)$ nilpotent orbits. 
We note that the relation (\ref{eq:rel1}) is satisfied if $M$ is a nilpotent matrix and if $N^{i_1i_2i_3i_4}N_{i_1i_2i_3i_4} = 0$. 
Since the operator $M$ satisfies the condition (\ref{eq:jordan}) and has maximal rank equal to one we can relate it to the minimal nilpotent orbit of $SU(8)$.

\subsection{$N=4$, $G_{global} = $ $SU(2) \times SU(8)$}
\begin{figure}[h!]
\center{
\begin{tikzpicture}[scale=0.7]

\draw (0.4,0) -- (0.8,0);
\draw[red, fill=red!20] (0,0) circle (0.4cm);
\draw (0,-0.7) node {4};

\draw (1.6,0) -- (2,0);
\draw[red, fill=red!20] (1.2,0) circle (0.4cm);
\draw (1.2,-0.7) node {3};

\draw (2.8,0) -- (3.2,0);
\draw[red, fill=red!20] (2.4,0) circle (0.4cm);
\draw (2.4,-0.7) node {2};

\draw[red, fill=red!20] (3.6,0) circle (0.4cm);
\draw (3.6,-0.7) node {1};

\draw (-0.4,0) -- (-0.8,0);
\draw[red, fill=red!20] (-1.2,0) circle (0.4cm);
\draw (-1.2,-0.7) node {3};

\draw (-1.6,0) -- (-2,0);

\draw (-2.8,0) -- (-3.2,0);
\draw[red, fill=red!20] (-2.4,0) circle (0.4cm);
\draw (-2.4,-0.7) node {2};

\draw[red, fill=red!20] (-3.6,0) circle (0.4cm);
\draw (-3.6,-0.7) node {1};

\draw (0,0.4) -- (0,0.8);
\draw (0,1.2) circle (0.4cm);
\draw (-0.65,1.2) node {2};

\draw (0,1.6) -- (0,2);
\draw[red, fill=red!20] (0,2.4) circle (0.4cm);
\draw (-0.65,2.4) node {1};

\end{tikzpicture}
}
\caption{Quiver diagram with $SU(2) \times SU(8)$ global symmetry group.\label{fig:su2su8}}
\end{figure}
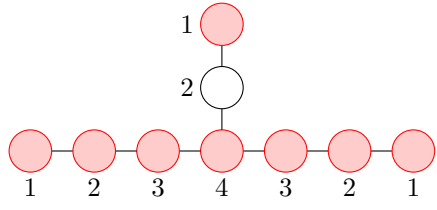

The quiver gauge theory with $SU(2) \times SU(8)$ global symmetry group is reported in figure \ref{fig:su2su8}.
The lowest orders of the expansion of its unrefined HS are
\begin{equation}
\label{eq:hssu2su8}
\begin{split}
 & \textrm{HS}_{SU(2) \times SU(8)}(t;1,...,1)_{SU(2) \times SU(8)} =  1 +66t^2 + 140t^3 +  2147t^4 +7588t^5 + 51247t^6 + o(t^6).
\end{split}
\end{equation}
We decompose the  previous HS under representations of $SU(2) \times SU(8)$ and we find the HWG
\begin{equation}
\label{eq:hwgsu2su8new}
\textrm{HWG}_{SU(2)\times SU(8)}(t;\nu,\mu_i)=\textrm{PE}[\nu^{2} t^2 +\mu_1\mu_7t^2 +\mu_4 \nu t^3 + t^4 + \mu_2\mu_6t^4 +\mu_4 \nu t^5+  \mu_3\mu_5t^6 + \mu_4^2t^8 - \nu^{2} \mu_4^2t^{10}].
\end{equation}
This result for the HWG follows the pattern outlined in equation (\ref{eq:hwgsu2nsu2}).
The first orders of the expansion of the corresponding HS read
\begin{equation}
\label{eq:hsu2su8}
\begin{split}
& \textrm{HS}_{SU(2) \times SU(8)}(t;x,y_i) = 1 + ([0;1,0,0,0,0,0,1]+[2;0,0,0,0,0,0,0])t^2 + [1;0,0,0,1,0,0,0]t^3 +\\
& ([0;0,0,0,0,0,0,0] + [0;0,1,0,0,0,1,0] + [0;2,0,0,0,0,0,2]+ [2;1,0,0,0,0,0,1]+ [4;0,0,0,0,0,0,0])t^4 +\\
& ([1;1,0,0,1,0,0,1]+ [3;0,0,0,1,0,0,0] + [1°;0,0,0,1,0,0,0])t^5 + ([0;0,0,0,1,0,0,0] + [2;0,0,0,2,0,0,0]+\\
& [0;1,0,0,0,0,0,1]+ [0;1,1,0,0,0,1,1]+ [0;3,0,0,0,0,0,3] + [2;0,0,0,0,0,0,0]+[2;0,1,0,0,0,1,0]+\\
& [2;2,0,0,0,0,0,2]+[4;1,0,0,0,0,0,1]+[6;0,0,0,0,0,0,0])t^6 + o(t^6) \ .
\end{split}\end{equation}
The first orders of the expansion of the PLog read
\begin{equation}
\label{eq:plogsu2su8}
\begin{split}
& \textrm{PLog}[\textrm{HS}_{SU(2) \times SU(8)}(t;x,y_i)] = ([2;0,0,0,0,0,0,0] + [0;1,0,0,0,0,0,1])t^2 + [1;0,0,0,1,0,0,0]t^3 \\
& - ([0;1,0,0,0,0,0,1]+[0;0,0,0,0,0,0,0])t^4 -([1;0,0,0,1,0,0,0] + [1;1,0,1,0,0,0,0] + \\ & [1;0,0,0,0,1,0,1])t^5 -([2;0,1,0,0,0,1,0] + [2;0,0,0,0,0,0,0]+ [0;0,0,1,0,1,0,0])t^6 +o(t^6).
\end{split}
\end{equation}

\subsubsection{The generators and their relations}
At the order $t^2$ of the expansion (\ref{eq:plogsu2su8}) there are two generators
\begin{align}
[0;1,0,0,0,0,0,1]:& \ \ \ M^{i_1}_{ \ \ i_2} \ \ \ \textrm{and} \ \ \ \textrm{Tr}[M]=0, \\
[2;0,0,0,0,0,0,0]:& \ \ \ C_{\alpha\beta}, 
\end{align}
where $i_1,i_2=1,...,8$ are $SU(8)$ indices while $\alpha,\beta=1,2$ are $SU(2)$ indices. At the order $t^3$ there is a further generator
\begin{align}
[1;0,0,0,1,0,0,0]:& \ \ \ N^{[i_1i_2i_3i_4]}_\alpha,
\end{align}
this generator transforms under the $\textbf{2} \times \textbf{70}$ representation of $SU(2) \times SU(8)$. At the order $t^4$ there are the relations
\begin{align}
[0;0,0,0,0,0,0,0] + [0;1,0,0,0,0,0,1] :& \ \ \ M^{i_1}_{\ \ j_1}M^{j_1}_{\ \ i_2} = c_1\delta^{i_1}_{\ \ i_2}C_{\alpha\beta}C^{\alpha\beta} \label{eq:relsu2su104}, \\
\end{align}
where $c_1 \in \mathbb{R}$. At the order $t^5$ there are the further relations\footnote{ For a generic tensor with the structure $P^{j[klm]}$ we define the projection to the irrep $\textbf{378}$ as
\begin{equation}
P^{ \{ j[klm] \} } = P^{j[klm]}-P^{[jklm]}.
\end{equation}
}

\begin{align}
[1;0,0,0,1,0,0,0]:& \ \ \ N^{i_1i_2i_3i_4}_{\alpha}C_{\beta\gamma}\epsilon^{\alpha\beta} + M^{[i_1}_{ \ \ j_1}N^{i_2i_3i_4]j_1}_{\alpha} = 0,\\
[1;0,0,0,0,1,0,1]:& \ \ \ M^{i_1}_{\ j_1}N^{j_1j_2j_3j_4}_{\alpha}\mid_{\textbf{2} \times \textbf{378}} = M^{\{ i_1}_{ \ \ \ j_1}N^{j_2j_3j_4 \} j_1}_{\alpha} = 0,\\
[1;1,0,1,0,0,0,0]:& \ \ \ M_{ \ \ i_1}^{j_1}N_{j_1j_2j_3j_4\alpha}\mid_{\textbf{2} \times \overline{\textbf{378}}} = M_{\ \ \{ i_1}^{j_1}N_{j_2j_3j_4 \} j_1 \alpha} = 0 \label{eq:rel2-378},
\end{align}
finally at the order $t^6$ there are the relations
\begin{align}
[2;0,0,0,0,0,0,0]:& \ \ \ N^{i_1i_2i_3i_4}_{(\alpha}N_{i_1i_2i_3i_4 \ \beta)} + C_{\alpha\beta}M^{i_1}_{\ \ j_1}M^{j_1}_{\ \ i_1} = 0, \\
[2;0,1,0,0,0,1,0]:& \ \ \ \left(N^{i_1i_2kl}_{(\alpha}N_{j_1j_2kl \ \beta)} + c_2C_{\alpha\beta}M^{[i_1}_{\ \ [j_1}M^{i_2]}_{ \ \ j_2]}\right)\mid_{\textbf{3} \times \textbf{720}}  = 0, \label{eq:relrank1} \\
[0;0,0,1,0,1,0,0]:& \ \ \ \left(N^{i_1i_2i_3l}_{(\alpha}N_{j_1j_2j_3l\beta)}\epsilon^{\alpha\beta} + c_3 M^{[i_1}_{\ \ [j_1}M^{i_2}_{\ \ j_2}M^{i_3]}_{\ \ j_3]}\right)\mid_{\textbf{1} \times \textbf{2352}}  = 0,   \label{eq:relrank}\\
\end{align}
where $c_1c_2,c_3 \in R$. In order to make contact with nilpotent orbits we set equal to zero all the generators with spin higher than 1.  Therefore we only consider the generator $M^{i_1}_{\ \ i_2}$ transforming under the adjoint representation of $SU(8)$ and the generator $C_{\alpha\beta}$ transforming in the adjoint of $SU(2)$.
The relations (\ref{eq:relsu2su104}) imply that the generators $M$ and $C$ are nilpotent and satisfy the condition (\ref{eq:jordan}).

\subsection{$N=5$, $G_{global} =SU(10)$}
\begin{figure}[h!]
\center{
\begin{tikzpicture}[scale=0.7]

\draw (0.4,0) -- (0.8,0);
\draw[red, fill=red!20] (0,0) circle (0.4cm);
\draw (0,-0.7) node {5};

\draw (1.6,0) -- (2,0);
\draw[red, fill=red!20] (1.2,0) circle (0.4cm);
\draw (1.2,-0.7) node {4};

\draw (2.8,0) -- (3.2,0);
\draw[red, fill=red!20] (2.4,0) circle (0.4cm);
\draw (2.4,-0.7) node {3};

\draw (4,0) -- (4.4,0);
\draw[red, fill=red!20] (3.6,0) circle (0.4cm);
\draw (3.6,-0.7) node {2};

\draw[red, fill=red!20] (4.8,0) circle (0.4cm);
\draw (4.8,-0.7) node {1};

\draw (-0.4,0) -- (-0.8,0);
\draw[red, fill=red!20] (-1.2,0) circle (0.4cm);
\draw (-1.2,-0.7) node {4};

\draw (-1.6,0) -- (-2,0);

\draw (-2.8,0) -- (-3.2,0);
\draw[red, fill=red!20] (-2.4,0) circle (0.4cm);
\draw (-2.4,-0.7) node {3};

\draw (-4,0) -- (-4.4,0);
\draw[red, fill=red!20] (-3.6,0) circle (0.4cm);
\draw (-3.6,-0.7) node {2};

\draw[red, fill=red!20] (-4.8,0) circle (0.4cm);
\draw (-4.8,-0.7) node {1};

\draw (0,0.4) -- (0,0.8);
\draw (0,1.2) circle (0.4cm);
\draw (-0.65,1.2) node {2};

\end{tikzpicture}
}
\caption{Quiver diagram with $SU(10)$ global symmetry group.\label{fig:su10}}
\end{figure}
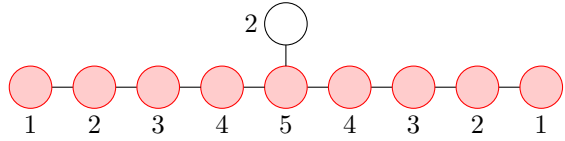

The quiver gauge theory with global symmetry group $SU(10)$ is reported in figure \ref{fig:su10}.
The first orders of the expansion of the corresponding unrefined Hilbert Series read 
\begin{equation}
\label{eq:hssu10new}
\begin{split}
& \textrm{HS}_{SU(10)}(t;1,...,1) =  1 +99t^2 + 252t^3 + 4851t^4 + 21252t^5 + 176352t^6 +o(t^{6}).
\end{split}
\end{equation}
We decompose the previous Hilbert Series under representation of  $SU(10)$. This way we find the HWG 
\begin{equation}
\label{eq:hwgsu10}
\textrm{HWG}_{SU(10)}(t;\mu_i)=\textrm{PE}[\mu_{1}\mu_{9}t^2 +\mu_5t^3 + (1+\mu_2\mu_8)t^4 + \mu_5t^5 +\mu_3\mu_7t^6 +\mu_4\mu_6t^8].
\end{equation}
This result for the HWG follows the pattern outlined in equation (\ref{eq:hwgsu2n}). The first orders of the expansion of the corresponding HS read
\begin{equation}\begin{split}
\label{eq:hsu10}
& \textrm{HS}_{SU(10)}(t,y_i)  = 1+ [1,0,0,0,0,0,0,0,1]t^2 + [0,0,0,0,1,0,0,0,0]t^3 + ([0,0,0,0,0,0,0,0,0]+\\
&[0,1,0,0,0,0,0,1,0]+[2,0,0,0,0,0,0,0,2])t^4 + ([0,0,0,0,1,0,0,0,0]+[1,0,0,0,1,0,0,0,1])t^5 \\
& ([0,0,1,0,0,0,1,0,0]+[0,0,0,0,2,0,0,0,0]+[1,0,0,0,0,0,0,0,1]+[1,1,0,0,0,0,0,1,1]\\
& [3,0,0,0,0,0,0,0,3])t^6 + o(t^6).
\end{split}\end{equation}
The first orders of the expansion of the PLog read
\begin{equation}
\label{eq:plogsu10}
\begin{split}
& \textrm{PLog}[\textrm{HS}_{SU(10)}(t;y_i)] = [1,0,0,0,0,0,0,0,1]t^2 + [0,0,0,0,1,0,0,0,0]t^3 - [1,0,0,0,0,0,0,0,1]t^4 + \\
& -([1,0,0,1,0,0,0,0,0] + [0,0,0,0,0,1,0,0,1])t^5 -[0,0,1,0,0,0,1,0,0]t^6 + o(t^6).
\end{split}
\end{equation}

\subsubsection{The generators and their relations}
At the order $t^2$ of the expansion (\ref{eq:plogsu10})
there is one generator
\begin{equation}
[1,0,0,0,0,0,0,0,1]: \ \ \ M^{i_1}_{\ \ i_2} \ \ \ \textrm{and} \ \ \ \textrm{Tr}[M]=0,
\end{equation}
where $i_1,i_2=1,...,10$ are $SU(10)$ indices. This operator transforms under the adjoint representation of $SU(10)$. At the order $t^3$ of the expansion there is a further generator
\begin{equation}
[0,0,0,0,1,0,0,0,0]: \ \ \ N^{[i_1i_2i_3i_4i_5]},
\end{equation} 
this operators transforms under the representation \textbf{252} of $SU(10)$. At the order $t^4$ there is a relation
\begin{equation}
[1,0,0,0,0,0,0,0,1]: \ \ \ M^{i_1}_{\ \ i_2}M^{i_2}_{\ \ i_3} - \frac{1}{10}\delta^{i_1}_{\ \ i_3}M^{i_4}_{\ \ i_5}M^{i_5}_{ \ \ i_4} = 0,
\end{equation}
this relation implies that $M$ is a nilpotent operator.
At the order $t^5$ there are two relations\footnote{For a tensor with  the structure $P^{i[jklp]}$ we define the projection to the irreducible representation \textbf{1848} as
\begin{equation}
P^{\{i[jklp]\}} = P^{i[jklp]} - P^{[ijklp]}.
\end{equation}
}
\begin{align}
[0,0,0,0,0,1,0,0,1]:& \ \ \ M^{i_1}_{\ \ j_1}N^{j_1i_2i_3i_4i_5}\mid_{\textbf{1848}} = M^{\{i_1}_{\ \ \ j_1}N^{i_2i_3i_4i_5 \} j_1} = 0,\label{eq:1848}\\
[1,0,0,1,0,0,0,0,0]:& \ \ \ M^{j_1}_{\ \ i_1}N_{j_1i_2i_3i_4i_5}\mid_{\overline{\textbf{1848}}} = M^{j_1}_{\ \ \{i_1}N_{i_2i_3i_4i_5 \} j_1} = 0,\label{eq:1848bar} 
\end{align}
while the relation $\overline{\textbf{1848}}$ is given by the conjugate of the  relation (\ref{eq:1848}). At the order $t^6$ there is the relation
\begin{equation}
\label{eq:rel10}
[0,0,1,0,0,0,1,0,0]: \ \ \ (M_{\ \ [i_1}^{  [j_1}M_{\ \ i_2}^{ j_2}M_{\ \ i_3]}^{ j_3]}+ N^{j_1j_2j_3i_4i_5}N_{i_1i_2i_3i_4i_5})|_{\textbf{12375}} = 0.
\end{equation}
In order to make contact with nilpotent orbits we set to zero the generator $N^{[i_1i_2i_3i_4i_5]}$, which has spin greater than one. Therefore $M$ is the only nilpotent generator of spin 1.

\subsection{$N=5$, $G_{global} = SU(2)\times SU(10)$}
\begin{figure}[h!]
\center{
\begin{tikzpicture}[scale=0.7]

\draw (0.4,0) -- (0.8,0);
\draw[red, fill=red!20] (0,0) circle (0.4cm);
\draw (0,-0.7) node {5};

\draw (1.6,0) -- (2,0);
\draw[red, fill=red!20] (1.2,0) circle (0.4cm);
\draw (1.2,-0.7) node {4};

\draw (2.8,0) -- (3.2,0);
\draw[red, fill=red!20] (2.4,0) circle (0.4cm);
\draw (2.4,-0.7) node {3};

\draw (4,0) -- (4.4,0);
\draw[red, fill=red!20] (3.6,0) circle (0.4cm);
\draw (3.6,-0.7) node {2};

\draw[red, fill=red!20] (4.8,0) circle (0.4cm);
\draw (4.8,-0.7) node {1};

\draw (-0.4,0) -- (-0.8,0);
\draw[red, fill=red!20] (-1.2,0) circle (0.4cm);
\draw (-1.2,-0.7) node {4};

\draw (-1.6,0) -- (-2,0);

\draw (-2.8,0) -- (-3.2,0);
\draw[red, fill=red!20] (-2.4,0) circle (0.4cm);
\draw (-2.4,-0.7) node {3};

\draw (-4,0) -- (-4.4,0);
\draw[red, fill=red!20] (-3.6,0) circle (0.4cm);
\draw (-3.6,-0.7) node {2};

\draw[red, fill=red!20] (-4.8,0) circle (0.4cm);
\draw (-4.8,-0.7) node {1};

\draw (0,0.4) -- (0,0.8);
\draw (0,1.2) circle (0.4cm);
\draw (-0.65,1.2) node {2};

\draw (0,1.6) -- (0,2.0);
\draw[red, fill=red!20] (0,2.4) circle (0.4cm);
\draw (-0.65,2.4) node {1};

\end{tikzpicture}
}
\caption{Quiver diagram with global symmetry group $SU(2)\times SU(10)$. \label{fig:su10su2}}
\end{figure}
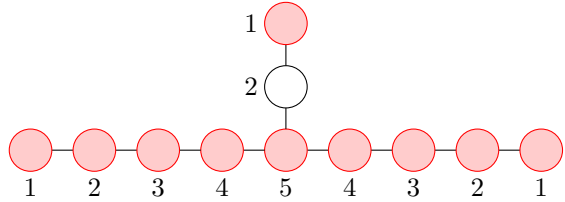

The quiver gauge theory with $SU(2) \times SU(10)$ global symmetry group is reported in figure \ref{fig:su10su2}.
The first orders of the expansion of the unrefined HS are
\begin{equation}
\label{eq:hssu2su10}
\begin{split}
& \textrm{HS}_{SU(2) \times SU(10)}(t;1,...1) =  1 + 102t^2 +5657t^4 + 215515t^6  +o(t^{6}).
\end{split}
\end{equation}
We decompose the Hilbert Series under representation of  $SU(2) \times 	SU(10)$. This way we find the HWG
\begin{equation}
\label{eq:hwgsu2su10}
\textrm{HWG}_{SU(2) \times SU(10)}(t;\nu,\mu_i)=\textrm{PE}[t^2(\mu_1\mu_9 + \nu^2)+ t^4(1 +\mu_2\mu_8 + \nu \mu_5)+ t^6(\nu \mu_5 +\mu_3\mu_7)+ t^8\mu_4\mu_6 + t^{10}\mu_5^2 - t^{12}\nu^2\mu_5^2].
\end{equation}
This result for the HWG follows the pattern outlined in equation (\ref{eq:hwgsu2nsu2}).
The first orders of the expansion of the Hilbert Series under  $SU(2) \times SU(10)$ representations read
\begin{equation}
\label{eq:hsu2su10}
\begin{split}
& \textrm{HS}_{SU(2)\times SU(10)}(t;x,y_i)  = 1 + ([0;1,0,0,0,0,0,0,0,1]+[2;0,0,0,0,0,0,0,0,0])t^2 +\\
& + ([0;0,0,0,0,0,0,0,0,0] + [1;0,0,0,0,1,0,0,0,0] + [0;0,1,0,0,0,0,0,1,0] +[4;0,0,0,0,0,0,0,0,0]\\
&  +[2;1,0,0,0,0,0,0,0,1] + [0;2,0,0,0,0,0,0,0,2] )t^4 + ([0;0,0,1,0,0,0,1,0,0]+[2;0,0,0,0,0,0,0,0,0]\\
& [3;0,0,0,0,1,0,0,0,0] + [2;0,1,0,0,0,0,0,1,0]+[6;0,0,0,0,0,0,0,0,0]+[0;1,0,0,0,0,0,0,0,1] \\
& + [1;1,0,0,0,1,0,0,0,1]+ [0;1,1,0,0,0,0,0,1,1] + [4;1,0,0,0,0,0,0,0,1] + [2;2,0,0,0,0,0,0,0,2]\\
& [0;3,0,0,0,0,0,0,0,3]+[1;0,0,0,0,1,0,0,0,0])t^6 + o(t^6).
\end{split}
\end{equation}
The Plethystic logarithm of the Hilbert series reads
\begin{equation}
\label{eq:plogsu2su10}
\begin{split}
& \textrm{PLog}[\textrm{HS}_{SU(2) \times SU(10)}(t;x,y_i)] = ([0;1,0,0,0,0,0,0,0,1] + [2;0,0,0,0,0,0,0,0,0])t^2 +  ([1;0,0,0,0,1,0,0,0,0]-\\
& [0;1,0,0,0,0,0,0,0,1]-[0;0,0,0,0,0,0,0,0,0])t^4 +   ([0;1,0,0,0,0,0,0,0,1] - [1;0,0,0,0,1,0,0,0,0] -\\
& [1;1,0,0,1,0,0,0,0,0] -[1;0,0,0,0,0,1,0,0,1])t^6 +o(t^6).
\end{split}
\end{equation}

\subsubsection{The generators and their relations}
At the order $t^2$ of the expansion (\ref{eq:plogsu2su10}) there are two generators
\begin{align}
[0;1,0,0,0,0,0,0,0,1]:& \ \ \ M^{i_1}_{ \ \ i_2} \ \ \ \textrm{and} \ \ \ \textrm{Tr}[M]=0, \\
[2;0,0,0,0,0,0,0,0,0]:& \ \ \ C_{\alpha\beta}, 
\end{align}
where $i_1,i_2=1,...,10$ are $SU(10)$ indices while $\alpha,\beta=1,2$ are $SU(2)$ indices. The generator $M$ transforms under the adjoint representation of $SU(10)$, while the generator $C$ transforms under the adjoint representation of $SU(2)$. At the order $t^4$ there is the further generator
\begin{align}
[1;0,0,0,0,1,0,0,0,0]:& \ \ \ N^{[i_1i_2i_3i_4i_5]}_{\alpha}, 
\end{align}
this generator transforms under the $\textbf{2} \times \textbf{252}$ representation of $SU(2) \times SU(10)$. Moreover, at the same order of the expansion, there are the relations
\begin{align}
[0;0,0,0,0,0,0,0,0,0] +[0;1,0,0,0,0,0,0,0,1] :& \ \ \ M^{i_1}_{\ \ j_1}M^{j_1}_{\ \ i_2} = c_1\delta^{i_1}_{\ \ i_2}C_{\alpha\beta}C^{\alpha\beta} \label{eq:rela},
\end{align}
where $c_1 \in \mathbb{R}$. At the order $t^6$  there are the following relations \footnote{For a generic tensor with the structure $P_{i[jklp]}$ we define the projection to the irrep \textbf{1848} as
\begin{equation}
P_{\{ijklp\}} = P_{i[jklp]} - P_{[ijklp]} \ .
\end{equation}
}

\begin{align}
[1;0,0,0,0,1,0,0,0,0]:& \ \ \
c_2M^{[i_1}_{\ \ j_1}N^{i_2i_3i_4i_5]j_1}_{\gamma} + N^{i_1i_2i_3i_4i_5}_{\alpha}C_{\beta\gamma}\epsilon^{\alpha\beta} = 0,\\
[1;0,0,0,0,0,1,0,0,1]:& \ \ \ M^{i_1}_{ \ \ j_1}N^{j_1j_2j_3j_4j_5}\mid_{\textbf{2} \times \textbf{1848}} =  M^{ \{ i_1}_{ \ \ \ j_1}N^{j_2j_3j_4j_5 \} j_1}_{\alpha} = 0,\\
[1;1,0,0,1,0,0,0,0,0]:& \ \ \ M_{\ \ i_1}^{j_1}N_{j_1j_2j_3j_4j_5}\mid_{\textbf{2} \times \overline{\textbf{1848}}} =  M_{\ \ \{ i_1}^{j_1}N_{j_2j_3j_4j_5 \} j_1\alpha} = 0,
\end{align}
where $c_2 \in R$. In order to make contact with nilpotent orbits we set to zero the generators with spin higher than 1. Therefore we keep only the operator $M^{i_1}_{\ \ i_2}$ transforming under the adjoint representation of $SU(10)$ and the operator $C_{\alpha\beta}$ transforming under the adjoint representation of $SU(2)$. The relations (\ref{eq:rela}) implies that $M$ and $C$ are nilpotent operators that satisfy the condition (\ref{eq:jordan}).

\subsection{$N=6$,  $G_{gloabl} = SU(12)$}
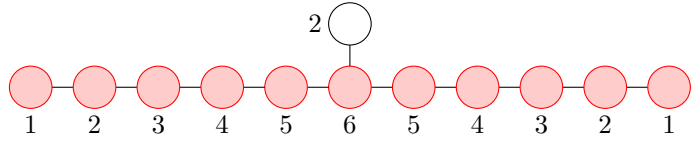
\begin{figure}[h!]
\center{
\begin{tikzpicture}[scale=0.7]

\draw (0.4,0) -- (0.8,0);
\draw[red, fill=red!20] (0,0) circle (0.4cm);
\draw (0,-0.7) node {6};

\draw (1.6,0) -- (2,0);
\draw[red, fill=red!20] (1.2,0) circle (0.4cm);
\draw (1.2,-0.7) node {5};

\draw (2.8,0) -- (3.2,0);
\draw[red, fill=red!20] (2.4,0) circle (0.4cm);
\draw (2.4,-0.7) node {4};

\draw (4,0) -- (4.4,0);
\draw[red, fill=red!20] (3.6,0) circle (0.4cm);
\draw (3.6,-0.7) node {3};

\draw (5.2,0)--(5.6,0);
\draw[red, fill=red!20] (4.8,0) circle (0.4cm);
\draw (4.8,-0.7) node {2};

\draw[red, fill=red!20] (6.0,0) circle (0.4cm);
\draw (6.0,-0.7) node {1};

\draw (-0.4,0) -- (-0.8,0);
\draw[red, fill=red!20] (-1.2,0) circle (0.4cm);
\draw (-1.2,-0.7) node {5};

\draw (-1.6,0) -- (-2,0);

\draw (-2.8,0) -- (-3.2,0);
\draw[red, fill=red!20] (-2.4,0) circle (0.4cm);
\draw (-2.4,-0.7) node {4};

\draw (-4,0) -- (-4.4,0);
\draw[red, fill=red!20] (-3.6,0) circle (0.4cm);
\draw (-3.6,-0.7) node {3};

\draw (-5.2,0) -- (-5.6,0);
\draw[red, fill=red!20] (-4.8,0) circle (0.4cm);
\draw (-4.8,-0.7) node {2};

\draw[red, fill=red!20] (-6,0) circle (0.4cm);
\draw (-6,-0.7) node {1};

\draw (0,0.4) -- (0,0.8);
\draw (0,1.2) circle (0.4cm);
\draw (-0.65,1.2) node {2};

\end{tikzpicture}
}
\caption{Quiver diagram with global symmetry group $SU(12)$ \label{fig:su12}.}
\end{figure}

The quiver gauge theory with $SU(12)$ global symmetry group is reported in figure \ref{fig:su12}.
The first orders of the expansion of the unrefined Hilbert Series read
\begin{equation}
\label{eq:hsu12}
\textrm{HS}_{SU(12)}(t;1,...1) = 1 + 143t^2 + 11077t^4 +592306t^6 + o(t^{6}).
\end{equation}
We decompose the  Hilbert Series under representation of  $SU(12)$. This way we get the HWG
\begin{equation}
\label{eq:HWGsu12}
\textrm{HWG}_{SU(12)}(t;\mu_i)=\textrm{PE}[\mu_{1}\mu_{11}t^2 +(1+\mu_2\mu_{10}+\mu_6)t^4 +(\mu_3\mu_9 +\mu_6)t^6 + \mu_4\mu_8t^8 + \mu_5\mu_7t^{10}].
\end{equation}
This result for the HWG follows the pattern outlined in equation (\ref{eq:hwgsu2n}).
The first orders of the expansion of the Hilbert Series in terms of $SU(12)$ are
\begin{equation}\begin{split}
\label{eq:exphs12}
& \textrm{HS}_{SU(12)}(t;y_i)  = 1 + [1,0,0,0,0,0,0,0,0,0,1]t^2 + ([0,0,0,0,0,0,0,0,0,0,0]\\
& [0,0,0,0,0,1,0,0,0,0,0] + [0,1,0,0,0,0,0,0,0,1,0] + [2,0,0,0,0,0,0,0,0,0,2])t^4+ \\
& ([0,0,0,0,0,1,0,0,0,0,0] + [0,0,1,0,0,0,0,0,1,0,0] + [1,0,0,0,0,0,0,0,0,0,1]\\
& +[1,0,0,0,0,1,0,0,0,0,1]+
 [1,1,0,0,0,0,0,0,0,1,1]+[3,0,0,0,0,0,0,0,0,0,3])t^6.
\end{split}\end{equation}
The first orders of the expansion of the PLog read
\begin{equation}
\label{eq:exppesu12}
\begin{split}
& \textrm{PLog}[\textrm{HS}_{SU(12)}(t;y_i)] = [1,0,0,0,0,0,0,0,0,0,1]t^2 + ([0,0,0,0,0,1,0,0,0,0,0]-[1,0,0,0,0,0,0,0,0,0,1])t^4 + \\
& ([1,0,0,0,0,0,0,0,0,0,1]-[1,0,0,0,1,0,0,0,0,0,0] -[0,0,0,0,0,0,1,0,0,0,1])t^6 + o(t^6).
\end{split}
\end{equation}

\subsubsection{The generators and their relations}
At the order $t^2$ there is the generator
\begin{equation}
[1,0,0,0,0,0,0,0,0,0,1]: \ \ \ M^{i_1}_{\ \ i_2} \ \ \ \textrm{and} \ \ \ \textrm{Tr}[M]=0,
\end{equation}
where $i_1,i_2=1,...,12$ are $SU(12)$ indices. This operator transforms under the adjoint representation of $SU(12)$. At the order $t^4$ there is another generator
\begin{equation}
[0,0,0,0,0,1,0,0,0,0,0]: \ \ \ N^{[i_1i_2i_3i_4i_5i_6]},
\end{equation}
this operator transforms under the representation \textbf{924} of $SU(12)$. Moreover there is the relation
\begin{equation}
[1,0,0,0,0,0,0,0,0,0,1]: \ \ \ M^{i_1}_{\  \ i_2}M^{i_2}_{\ \ i_3} - \frac{1}{12}\delta^{i_1}_{\ \ i_3}M^{i_4}_{\ \ i_5}M^{i_5}_{\ \ i_4}= 0,
\end{equation}
therefore $M$ is a nilpotent operator. At the order $t^6$
there are two relations \footnote{for a generic tensor with the structure $P^{i[jklpq]}$ we introduce the projection to the irreducible representation \textbf{8580} as
\begin{equation}
P^{\{ijklpq\}} = P^{i[jklpq]} - P^{[ijklpq]} \ .
\end{equation}
}
\begin{align}
[0,0,0,0,0,0,1,0,0,0,1]:& \ \ \ M^{i_1}_{\ \ j_1}N^{i_2i_3i_4i_5i_6 j_1}\mid_{\textbf{8580}} = M^{\{i_1}_{\ \ \ j_1}N^{i_2i_3i_4i_5i_6 \} j_1} = 0, \label{eq:8580} \\
[1,0,0,0,1,0,0,0,0,0,0]:& \ \ \ M_{\ \ i_1}^{j_1}N_{i_2i_3i_4i_5i_6 j_1}\mid_{\overline{\textbf{8580}}} = M_{\ \ \{i_1}^{j_1}N_{i_2i_3i_4i_5i_6 \} j_1} = 0, \label{eq:8580bar}
\end{align}
in order to make contact with nilpotent orbits we set to zero all the generators with spin higher than one. Therefore we keep only the nilpotent generator $M^{i_1}_{\ \ i_2}$.

\subsection{$N=6$, $G_{gloabl} = SU(2) \times SU(12) $}
\begin{figure}[h!]
\center{
\begin{tikzpicture}[scale=0.7]

\draw (0.4,0) -- (0.8,0);
\draw[red, fill=red!20] (0,0) circle (0.4cm);
\draw (0,-0.7) node {6};

\draw (1.6,0) -- (2,0);
\draw[red, fill=red!20] (1.2,0) circle (0.4cm);
\draw (1.2,-0.7) node {5};

\draw (2.8,0) -- (3.2,0);
\draw[red, fill=red!20] (2.4,0) circle (0.4cm);
\draw (2.4,-0.7) node {4};

\draw (4,0) -- (4.4,0);
\draw[red, fill=red!20] (3.6,0) circle (0.4cm);
\draw (3.6,-0.7) node {3};

\draw (5.2,0)--(5.6,0);
\draw[red, fill=red!20] (4.8,0) circle (0.4cm);
\draw (4.8,-0.7) node {2};

\draw[red, fill=red!20] (6.0,0) circle (0.4cm);
\draw (6.0,-0.7) node {1};

\draw (-0.4,0) -- (-0.8,0);
\draw[red, fill=red!20] (-1.2,0) circle (0.4cm);
\draw (-1.2,-0.7) node {5};

\draw (-1.6,0) -- (-2,0);

\draw (-2.8,0) -- (-3.2,0);
\draw[red, fill=red!20] (-2.4,0) circle (0.4cm);
\draw (-2.4,-0.7) node {4};

\draw (-4,0) -- (-4.4,0);
\draw[red, fill=red!20] (-3.6,0) circle (0.4cm);
\draw (-3.6,-0.7) node {3};

\draw (-5.2,0) -- (-5.6,0);
\draw[red, fill=red!20] (-4.8,0) circle (0.4cm);
\draw (-4.8,-0.7) node {2};

\draw[red, fill=red!20] (-6,0) circle (0.4cm);
\draw (-6,-0.7) node {1};

\draw (0,0.4) -- (0,0.8);
\draw (0,1.2) circle (0.4cm);
\draw (-0.65,1.2) node {2};

\draw (0,1.6) -- (0,2.0);
\draw[red, fill=red!20] (0,2.4) circle (0.4cm);
\draw (-0.65,2.4) node {1};

\end{tikzpicture}
}
\caption{Quiver diagram with global symmetry group $SU(2)\times  SU(12)$ \label{fig:su12su2}}
\end{figure}
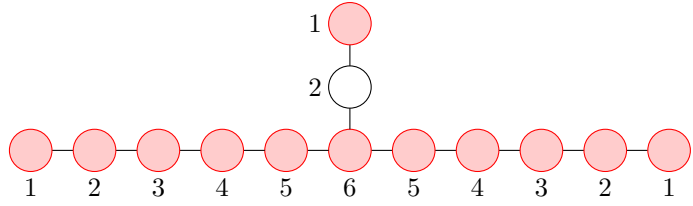

The quiver gauge theory with global symmetry group $SU(2) \times SU(12)$ is reported in figure \ref{fig:su12su2}.
The first orders of the expansion of the corresponding unrefined Hilbert Series read
\begin{equation}
\label{eq:hsu2su12}
\begin{split}
& \textrm{HS}_{SU(2) \times SU(12)}(t;1,...,1) = 1 + 146t^2 + 10587t^4 +1848t^5 + 508515t^6 + o(t^6).
\end{split}
\end{equation}
We decompose  the Hilbert Series under representations of  $SU(2) \times SU(12)$. This way we get the HWG
\begin{equation}
\label{eq:hwgsu2su12}
\begin{split}
& \textrm{HWG}_{SU(2) \times SU(12)}(t;\nu,\mu_i)=
\textrm{PE}[(\nu^2+\mu_{1}\mu_{11})t^2 +(1+\mu_2\mu_{10})t^4 +\nu \mu_6t^5+ \mu_3\mu_9t^6 + \nu \mu_6t^7 + \mu_4\mu_8t^8 + \\
& \mu_5\mu_7t^{10} +\mu_6^2t^{12} -\nu^2\mu_6^2t^{14}].
\end{split}
\end{equation}
This result for the HWG follows the pattern outlined in equation (\ref{eq:hwgsu2nsu2}). The first orders of the expansion of the Hilbert Series in terms of $SU(2) \times SU(12) $ representations are
\begin{equation}
\label{eq:hsu2su12expansion}
\begin{split}
& \textrm{HS}_{SU(2) \times SU(12)}(t;x,y_i)  = 1+ ([0;1,0,0,0,0,0,0,0,0,0,1] + [2;0,0,0,0,0,0,0,0,0,0,0])t^2 \\\
& ([0;0,0,0,0,0,0,0,0,0,0,0]+[0;0,1,0,0,0,0,0,0,0,1,0]+[0;2,0,0,0,0,0,0,0,0,0,2]+\\
& [2;1,0,0,0,0,0,0,0,0,0,1]+[4;0,0,0,0,0,0,0,0,0,0,0])t^4 + [1;0,0,0,0,0,1,0,0,0,0,0]t^5 +\\
& ([0;0,0,1,0,0,0,0,0,1,0,0]+ [0;1,0,0,0,0,0,0,0,0,0,1] + [0;1,1,0,0,0,0,0,0,0,1,1]+ \\
& [0;3,0,0,0,0,0,0,0,0,0,3]+ [2;0,0,0,0,0,0,0,0,0,0,0]+[2;0,1,0,0,0,0,0,0,0,1,0]\\
& [2;2,0,0,0,0,0,0,0,0,0,2]+ [4;1,0,0,0,0,0,0,0,0,0,1]+[6;0,0,0,0,0,0,0,0,0,0,0] )t^6 +o(t^6).
\end{split}
\end{equation}
The PLog of the corresponding HS reads
\begin{equation}
\label{eq:plogsu2su12}
\begin{split}
& \textrm{PLog}[\textrm{HS}_{SU(2) \times SU(12)}(t;x,y_i)]  =([0;1,0,0,0,0,0,0,0,0,0,0,0,1]+[2;0,0,0,0,0,0,0,0,0,0,0])t^2 + \\
& - ([0;1,0,0,0,0,0,0,0,0,0,1]+[0;0,0,0,0,0,0,0,0,0,0,0])t^4 +[1;0,0,0,0,0,1,0,0,0,0,0]t^5 +\\
&  [0;1,0,0,0,0,0,0,0,0,0,1]t^6 +o(t^6).
\end{split}\end{equation}

\subsubsection{The generators and their relations}
At the order $t^2$ of the expansion (\ref{eq:plogsu2su12}) there are two generators
\begin{align}
[0;1,0,0,0,0,0,0,0,0,0,1]:& \ \ \ M^{i_1}_{ \ \ i_2} \ \ \ \textrm{and} \ \ \ \textrm{Tr}[M]=0, \\
[0;0,0,0,0,0,0,0,0,0,0,0]:& \ \ \ C_{\alpha\beta}, \label{eq:rel4su12}
\end{align}
where $i_1,i_2=1,...,12$ are $SU(12)$ indices while $\alpha,\beta=1,2$ are $SU(2)$ indices. The generator $M$ transforms under the adjoint representation of $SU(12)$, while the operator $C$ transforms under the adjoint representation of $SU(2)$. At the order $t^4$ there are the relations
\begin{align}
[0;0,0,0,0,0,0,0,0,0,0,0] + [0;1,0,0,0,0,0,0,0,0,0,1] :& \ \ \ M^{i_1}_{\ \ j_1}M^{j_1}_{\ \ i_2} = c_1\delta^{i_1}_{\ \ i_2}C_{\alpha\beta}C^{\alpha\beta} \label{eq:relb},
\end{align}
where $c_1 \in R$. At the order $t^5$ there is a further generator
\begin{align}
[1;0,0,0,0,0,1,0,0,0,0,0]:& \ \ \ N^{[i_1i_2i_3i_4i_5i_6]}_{ \ \alpha}, \
\end{align}
this generator transforms under the $\textbf{2} \times \textbf{924}$ representation of $SU(2) \times SU(12)$.

We set equal to zero all the generators with spin higher than 1. This way we keep only the operator $M^{i_1}_{\ \ i_2}$ and the operator $C_{\alpha\beta}$. The relation (\ref{eq:relb}) implies that $M$ and $C$ are nilpotent and satisfy the condition (\ref{eq:jordan}).

\subsection{The mirror of the $(k)-[2N]$ theory}
\label{sec:(n)-[2n] theory}
We test the expression (\ref{eq:hwgk2n}) performing the computation of the HWG for the mirror of the $(k)-[2N]$ theory, with $k \leq N$. The relation between this class of theories and closure of nilpotent orbits has been extensively studied in \cite{Cabrera:2016vvv}. The result is that each theory can be parametrized by a $N \times N$ nilpotent matrix whose rank is at most equal to $k$. So that the Coulomb branch $\mathcal{M_{C}}$ of the mirror of the $(k)-[2N]$ can be written as the closure of the nilpotent orbit parametrized by the partition
$(2^{k},1^{2N-2k})$.
\begin{equation}
\mathcal{M_{C}} = \bar{\mathcal{O}}_{(2^{k},1^{2N-2k})} \ .
\end{equation}

\section{Theories with orthogonal global symmetry group}
\label{sec:globalort}
In this section we test the expression of the HWG (\ref{eq:hwgso4n6u1}) when $N=1,2$.
\subsection{$N=1,  \ G_{global}  = E_6 \supset SO(10) \times U(1)$}
The computation of the HS for this case is reported in section \ref{subsec:e6}.
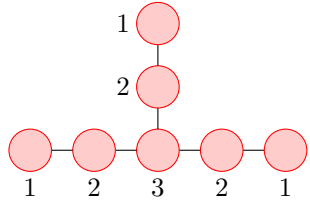
\begin{figure}[h!]
\center{
\begin{tikzpicture}[scale=0.7]

\draw (0.4,0) -- (0.8,0);
\draw[red, fill=red!20] (0,0) circle (0.4cm);
\draw (0,-0.7) node {3};

\draw (1.6,0) -- (2,0);
\draw[red, fill=red!20] (1.2,0) circle (0.4cm);
\draw (1.2,-0.7) node {2};

\draw[red, fill=red!20] (2.4,0) circle (0.4cm);
\draw (2.4,-0.7) node {1};

\draw (-0.4,0) -- (-0.8,0);
\draw[red, fill=red!20] (-1.2,0) circle (0.4cm);
\draw (-1.2,-0.7) node {2};

\draw (-1.6,0) -- (-2,0);
\draw[red, fill=red!20] (-2.4,0) circle (0.4cm);
\draw (-2.4,-0.7) node {1};

\draw (0,0.4) -- (0,0.8);
\draw[red, fill=red!20] (0,1.2) circle (0.4cm);
\draw (-0.65,1.2) node {2};

\draw (0,1.6) -- (0,2);
\draw[red, fill=red!20] (0,2.4) circle (0.4cm);
\draw (-0.65,2.4) node {1};

\end{tikzpicture}
}
\caption{Quiver diagram with $E_{6}$ global symmetry group. \label{fig:so10u1}}
\end{figure} However this time we decompose representations of $E_6$ under  representations of $SO(10) \times U(1)$. This way we find the HWG \footnote{ This HWG is an agreement with the result found in \cite{Cremonesi:2015lsa}.}   
\begin{equation}
\textrm{HWG}_{SO(10) \times U(1)}(t;q;\mu_i) = \textrm{PE}\left[t^2 +
\mu_2t^2 + \mu_4qt^2 + \frac{\mu_5}{q}t^2\right],
\end{equation}
where $q$ is the fugacity for the $U(1)$ charge and the various $\mu_i$ are the $SO(10)$ highest weights. The first orders of the expansion of the corresponding Hilbert Series are
\begin{equation}\begin{split}
& \textrm{HS}_{SO(10) \times U(1)}(t;y_i,q)= 1 + ([0,0,0,0,0] + [0,1,0,0,0]+ q^{-1}[0,0,0,0,1]+q[0,0,0,1,0])t^2 \\
& + ([0,0,0,0,0]+[0,1,0,0,0]+[0,2,0,0,0]+[0,0,0,1,1] + q^{-1}([0,1,0,0,1]+[0,0,0,0,1])+\\
& q([0,1,0,1,0]+[0,0,0,1,0])+q^{-2}[0,0,0,0,2]+q^2[0,0,0,2,0] )t^4 + o(t^4).
\end{split}\end{equation}
The first orders of the expansion of the corresponding Plethystic logarithm read
\begin{equation}\begin{split}
& \textrm{Plog}[\textrm{HS}_{SO(10)\times U(1)}(t;y_i,q)]= (1 + [0,1,0,0,0]+q^{-1}[0,0,0,0,1]+q[0,0,0,1,0])t^2 + \\
& -(2 + [0,1,0,0,0] + [2,0,0,0,0]+ [0,0,0,1,1]+ q([1,0,0,0,1]+[0,0,0,1,0]) +\\
& q^{-1}([1,0,0,1,0]+[0,0,0,0,1])+(q^2+q^{-2})[1,0,0,0,0])t^4 +o(t^4).
\end{split}\end{equation}
\subsubsection{The generators and their relations}
We use the same conventions employed in \cite{Cremonesi:2015lsa}. We denote $SO(10)$ vector indices with Latin letters $a,b,=1,...,10$, while we denote $SO(10)$ spinor indices with Greek letters $\alpha,\beta=1,...,16$ \footnote{The Kronecker delta has the following form
\begin{equation}
\delta^{\alpha}_{\beta},
\end{equation}
while the gamma matrices take the forms
\begin{equation}
(\gamma^{a})_{\alpha\beta} \ \  \textrm{and} \ \ (\gamma^{a})^{\alpha\beta}.
\end{equation} 
The product of two gamma matrices takes the form
\begin{equation}
(\gamma^{ab})^{\alpha}_{\tau} = (\gamma^{[a})^{\alpha\sigma}(\gamma^{b]})_{\sigma\tau},
\end{equation} 
while the product of four gamma matrices reads
\begin{equation}
(\gamma^{abcd})^{\alpha}_{\ \beta}= (\gamma^{[a})^{\alpha\tau_1}(\gamma^{b})_{\tau_1\tau_2}(\gamma^{c})^{\tau_2\tau_3}(\gamma^{d]})_{\tau_3\beta} \ .
\end{equation}
}. At the order $t^2$ there are the following four generators
\begin{equation}
M^{ab}, \ \ \ T^{\alpha}, \ \ \ \tilde{T}_{\alpha}, \ \ \ S,
\end{equation}
where $M^{ab}$ is a $10 \times 10$ antisymmetric matrix. The operator $T^{\alpha}$ transforms under the spinorial representation while $\tilde{T}^{\alpha}$ transforms under the complex conjugated
representation, finally $S$ is a scalar operator. At the order $t^4$ there are the following relations \footnote{A similar analysis of this moduli space has been performed in \cite{Cremonesi:2015lsa}.}
\begin{align}
[2,0,0,0,0]+[0,0,0,0,0]:& \ \ \ M^{ab}M^{bc}=(T^{\alpha}\tilde{T}_{\alpha})\delta^{ac}, \label{eq:mnil}\\
[0,0,0,1,1]:& \ \ \ M^{[a_1a_2}M^{a_3a_4]} = \tilde{T}_{\beta}(\gamma^{a_1...a_4})^{\beta}_{ \ \alpha}T^{\alpha}, \label{eq:mrank}\\
[0,0,0,0,0]:& \ \ \ S^2=T^{\alpha}\tilde{T}_{\alpha},\label{eq:snil} \\
[0,1,0,0,0]:& \ \ \ SM^{ab} = \tilde{T}_{\beta}(\gamma^{ab})^{\beta}_{\ \alpha}T^{\alpha}, \label{eq:rel13} \\
q([1,0,0,0,1]+[0,0,0,1,0]):& \ \ \ M^{ab}T^{\alpha}(\gamma^{b})^{\beta}_{\ \alpha}=ST^{\alpha}(\gamma^{a})^{\beta}_{\ \alpha}, \label{eq:rel11}\\
q^{-1}([1,0,0,1,0]+[0,0,0,0,1]):& \ \ \ M^{ab}\tilde{T}_{\beta}(\gamma^{b})^{\beta}_{\ \alpha} = S\tilde{T}_{\beta}(\gamma^{a})^{\beta}_{\ \alpha},\\
(q^2+q^{-2})[1,0,0,0,0]:& \ \ \ T^{\alpha}T^{\beta}(\gamma^{a})_{\alpha\beta}=\tilde{T}_{\alpha}\tilde{T}_{\beta}(\gamma^{a})^{\alpha\beta} = 0.\label{eq:rel12}
\end{align}

\subsection{N=2, \ $G_{global} = SO(14) \times U(1)$}
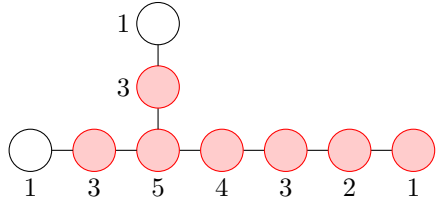
\begin{figure}[h]
\center{
\begin{tikzpicture}[scale=0.7]

\draw (0.4,0) -- (0.8,0);
\draw[red, fill=red!20] (0,0) circle (0.4cm);
\draw (0,-0.7) node {5};

\draw (1.6,0) -- (2,0);
\draw[red, fill=red!20] (1.2,0) circle (0.4cm);
\draw (1.2,-0.7) node {4};

\draw (2.8,0) -- (3.2,0);
\draw[red, fill=red!20] (2.4,0) circle (0.4cm);
\draw (2.4,-0.7) node {3};

\draw (4,0) -- (4.4,0);
\draw[red, fill=red!20] (3.6,0) circle (0.4cm);
\draw (3.6,-0.7) node {2};

\draw[red, fill=red!20] (4.8,0) circle (0.4cm);
\draw (4.8,-0.7) node {1};

\draw (-0.4,0) -- (-0.8,0);
\draw[red, fill=red!20] (-1.2,0) circle (0.4cm);
\draw (-1.2,-0.7) node {3};

\draw (-1.6,0) -- (-2,0);

\draw (-2.4,0) circle (0.4cm);
\draw (-2.4,-0.7) node {1};

\draw (0,0.4) -- (0,0.8);
\draw[red, fill=red!20] (0,1.2) circle (0.4cm);
\draw (-0.65,1.2) node {3};

\draw (0,1.6) -- (0,2.0);
\draw (0,2.4) circle (0.4cm);
\draw (-0.65,2.4) node {1};

\end{tikzpicture}
}
\caption{Quiver diagram with global symmetry group $SO(14)\times U(1)$.\label{fig:so14u1}}
\end{figure}

The quiver diagram with $SO(14)\times U(1)$ global symmetry is reported in figure \ref{fig:so14u1}.
The first orders of the expansion of the unrefined HS are
\begin{equation}\begin{split}
& \textrm{HS}_{SO(14) \times U(1)}(t;1,...1)= 1+92 t^2+128 t^3+4173 t^4+9984 t^5+127920 t^6+ o(t^6).
\end{split}\end{equation}
The corresponding HWG reads
\begin{equation}
\label{eq:hwgso14u1}
\textrm{HWG}_{SO(14) \times U(1)}(t;\mu_i,q)=\textrm{PE}\left[t^2 + \mu_2t^2  +  q \mu_6t^3 +\frac{\mu_7}{q}t^3 +   \mu_4t^4 \right],
\end{equation}
where the $\mu_i$ are $SO(14)$ highest weights while $q$ is the $U(1)$ fugacity. The HS expressed in terms of $SO(14) \times U(1)$ representations reads
\begin{equation}\begin{split}
& \textrm{HS}(t;y_i,q)_{SO(14)\times U(1)}= 1 + [0,1,0,0,0,0,0]t^2 + (q^{-1}[0,0,0,0,0,0,1]+q[0,0,0,0,0,1,0])t^3\\
& + (1+[0,0,0,1,0,0,0]+[0,2,0,0,0,0,0]+[0,1,0,0,0,0,0])t^4 + (q^{-1}[0,0,0,0,0,0,1]+\\
& q^{-1}[0,1,0,0,0,0,1]+ q[0,0,0,0,0,1,0]+q[0,1,0,0,0,1,0])t^5 + (1+[0,0,0,0,0,1,1]+\\
& [0,0,0,1,0,0,0]+ [0,1,0,0,0,0,0]+[0,1,0,1,0,0,0]+[0,2,0,0,0,0,0]+[0,3,0,0,0,0,0]\\
& + q^2[0,0,0,0,0,2,0]+q^{-2}[0,0,0,0,0,0,2])t^6 + o(t^6).
\end{split}\end{equation}
The corresponding Plethystic logarithm reads
\begin{equation}\begin{split}
& \textrm{PLog}[\textrm{HS}_{SO(14) \times U(1)}(t;y_i,q)] = ([0,1,0,0,0,0,0] + [0,0,0,0,0,0,0])t^2 + \\
& + (q[0,0,0,0,0,1,0]+q^{-1}[0,0,0,0,0,0,1])t^3 - ([0,0,0,0,0,0,0] + [2,0,0,0,0,0,0])t^4+ \\
& -( q([1,0,0,0,0,0,1]+[0,0,0,0,0,1,0]) + q^{-1}([1,0,0,0,0,1,0]+[0,0,0,0,0,0,1]))t^5 +\\
& -(1+[0,0,0,0,0,1,1]+[0,0,0,1,0,0,0]+[0,1,0,0,0,0,0]-[2,0,0,0,0,0,0] \\
& +(q^2+q^{-2})[0,0,1,0,0,0,0])t^6 + o(t^6).
\end{split}\end{equation}

\subsubsection{The generators and their relations}
At the order $t^2$ there are two generators
\begin{equation}
M^{ab} \ \ \ \textrm{and} \ \ \ S,
\end{equation}
where $M^{ab}$ is an antisymmetric matrix, while $S$ is a scalar operator. At the order $t^3$ there two further generators
\begin{equation}
T^{\alpha} \ \ \ \textrm{and} \ \ \ \tilde{T}_{\alpha},
\end{equation}	
the operator $T^{\alpha}$ transforms under the spinorial representation of $SO(14)$, while the operator $\tilde{T}_{\alpha}$ transforms under the complex conjugate representation. At the order $t^4$  there are the relations
\begin{equation}
[2,0,0,0,0,0,0] + [0,0,0,0,0,0,0]: \ \ \ M^{ab}M^{bc} = S^2\delta^{ac}. \label{eq:relnilm}
\end{equation} At the order $t^5$ we have the further relations
\begin{align}
q([1,0,0,0,0,0,1]+[0,0,0,0,0,1,0]):& \ \ \ \ \ M^{ab}T^{\alpha}(\gamma^{b})^{\beta}_{\ \alpha} = ST^{\alpha}(\gamma^{a})^{\beta}_{\ \alpha},\\
q^{-1}([1,0,0,0,0,1,0]+[0,0,0,0,0,0,1]):& \ \ \ \ \ M^{ab}\tilde{T}_{\beta}(\gamma^{b})^{\beta}_{\ \alpha} = S\tilde{T}_{\beta}(\gamma^{a})^{\beta}_{\ \alpha}.
\end{align}
At the order $t^6$ there are the relations
\begin{align}
[0,0,0,0,0,0,0]:& \ \ \ \ \ S^3=T^{\alpha}\tilde{T}_{\alpha}, \label{eq:relnils}\\
[0,0,0,0,0,1,1]:& \ \ \ \ \ M^{[a_1a_2}M^{a_3a_4}M^{a_5a_6]}=\tilde{T}_{\beta}(\gamma^{a_1...a_6})^{\beta}_{\ \alpha}T^{\alpha}, \label{rel:cm} \\
[0,1,0,0,0,0,0]:& \ \ \ \ \ SM^{ab}=\tilde{T}_{\beta}(\gamma^{ab})^{\beta}_{\ \alpha}T^{\alpha}, \label{rel:int} \\
(q^2+q^{-2})[0,0,1,0,0,0,0]:& \ \ \ \ \ T^{\alpha}T^{\beta}(\gamma^{abc})_{\alpha\beta}=\tilde{T}_{\alpha}\tilde{T}_{\beta}(\gamma^{abc})^{\alpha\beta} = 0 \ . 
\end{align}

\subsection{$N=3$ \ $G_{global}= E_8 \supset SO(16)$ }
\begin{figure}[h]
\center{
\begin{tikzpicture}[scale=0.70]

\draw (0.4,0) -- (0.8,0);
\draw[red, fill=red!20] (0,0) circle (0.4cm);
\draw (0,-0.85) node {6};

\draw (1.6,0) -- (2,0);
\draw[red, fill=red!20] (1.2,0) circle (0.4cm);
\draw (1.2,-0.85) node {5};

\draw (2.8,0) -- (3.2,0);
\draw[red, fill=red!20] (2.4,0) circle (0.4cm);
\draw (2.4,-0.85) node {4};

\draw (4,0) -- (4.4,0);
\draw[red, fill=red!20] (3.6,0) circle (0.4cm);
\draw (3.6,-0.85) node {3};

\draw (5.2,0) -- (5.6,0);
\draw[red, fill=red!20] (4.8,0) circle (0.4cm);
\draw (4.8,-0.85) node {2};

\draw[red, fill=red!20] (6,0) circle (0.4cm);
\draw (6,-0.85) node {1};

\draw (-0.4,0) -- (-0.8,0);
\draw[red,fill=red!20] (-1.2,0) circle (0.4cm);
\draw (-1.2,-0.85) node {4};

\draw (-1.6,0) -- (-2,0);
\draw[red,fill=red!20] (-2.4,0) circle (0.4cm);
\draw (-2.4,-0.85) node {2};

\draw (0,0.4) -- (0,0.8);
\draw[red,fill=red!20] (0,1.2) circle (0.4cm);
\draw (-0.85,1.2) node {3};

\end{tikzpicture}
}
\caption{quiver diagram with $E_8$ global symmetry group. \label{fig:e8}}
\end{figure}
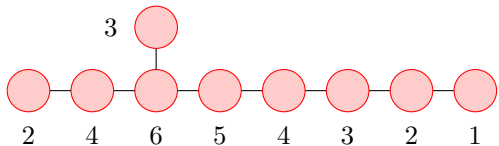

The quiver gauge theory with $E_8$ global symmetry group is reported in figure \ref{fig:e8}.

The first orders of the expansion of the unrefined Hilbert Series are
\begin{equation}
\textrm{HS}_{SO(16)}(t) = 1+ 248t^2 + 27000t^4 + 1763125t^6 + o(t^6).
\end{equation}
The corresponding HWG reads
\begin{equation}
\textrm{HWG}_{SO(16)}(t;\mu_i) = \textrm{PE}\left[ (\mu_2+\mu_8)t^2 + (1+\mu_4+\mu_8)t^4 +\mu_6t^6 \right],
\end{equation}
where the $\mu_i$ are $SO(16)$ highest weights. The HS expressed in terms of $SO(16)$ representations reads
\begin{equation}\begin{split}
\label{eq:hss016}
& \textrm{HS}_{SO(16)}(t;\mu_i)= 1 + ([0,1,0,0,0,0,0,0]+[0,0,0,0,0,0,0,1])t^2 + (1+ [0,0,0,1,0,0,0,0] + \\
& [0,0,0,0,0,0,0,1]+[0,2,0,0,0,0,0,0] +[0,0,0,0,0,0,0,2] + [0,1,0,0,0,0,0,1] )t^4 + o(t^4).
\end{split}\end{equation}
The first orders of the expansion of the corresponding Plethystic logarithm are
\begin{equation}\begin{split}
\label{eq:plogs016}
& \textrm{PLog}[\textrm{HS}_{SO(16)}(t;\mu_i)]=([0,0,0,0,0,0,0,1]+[0,1,0,0,0,0,0,0])t^2
\\
& - (1+ [0,0,0,1,0,0,0,0] + [1,0,0,0,0,0,1,0]+[2,0,0,0,0,0,0,0])t^4 + o(t^4),
\end{split}\end{equation}

\subsubsection{The generators and their relations}
At the order $t^2$ there two generators
\begin{equation}
X^{ab} \ \ \ \textrm{and} \ \ \ \tilde{T}_{\alpha},
\end{equation} 
where we use $a,b=1,...,16$ to denote $SO(16)$ vector indices and $\alpha,\beta = 1,...,128$ to denote $SO(16)$ spinor indices. The operator $X^{ab}$ is an antisymmetric matrix in the adjoint representation of $SO(16)$. The operator $\tilde{T}_{\alpha}$ transforms under the conjugate spinor representation of $SO(16)$. At the order $t^4$ there are the relations
\begin{align}
1+[2,0,0,0,0,0,0,0]:& \ \ \ \ \ X^2=0, \label{rel:nilpx}\\
[1,0,0,0,0,0,1,0]:& \ \ \ \ \ X^{ab}\tilde{T}_{\beta}(\gamma^{b})^{\beta}_{\ \alpha}=0, \\
[0,0,0,1,0,0,0,0]: & \ \ \ \ \ \tilde{T}_{\alpha}\tilde{T}_{\beta}(\gamma^{abcd})^{\alpha \beta} =0.
\end{align}
The relation (\ref{rel:nilpx}) tells us that $X$ is a nilpotent operator. 

\section{Conclusions}
\label{sec:conclusions}

In this paper we found and we successfully tested the expressions of the HWG and of the corresponding HS for the Coulomb branch of the mirror of some families of $3d$ Sicilian theories. In particular we analysed the case of theories with unitary and orthogonal global symmetry group. In all the cases we decomposed the corresponding HWG and HS under representation of $G_{global}$. Moreover we explicitly checked that the numerator of the unrefined Hilbert Series is given by a palindromic polynomial and we studied the generators and the relations arising from the power series expansion of the corresponding Plethystic Logarithm. It would be interesting to extend the previous analysis also to different theories which exhibit a more involved global symmetry group. We postpone such study for future work.

\section*{Acknowledgements}
 A.P. would like to acknowledge the String Theory Group of the Queen Mary University of London and  the String Theory Group of the Imperial College of London for very kind hospitality during the initial part of this project. Moreover the authors would like to acknowledge the support of the COST Action MP1210 STSM. A. H. and A.P. gratefully acknowledge hospitality at the Simons Center for Geometry and Physics, Stony Brook University where the last part of the research for this paper was performed. Furthermore the authors would like to acknowledge Diego Rodríguez-Gómez, Davíd Rodríguez-Fernandez, Andres Viña Escalar, Rudolph Kalveks, Giulia Ferlito, Antoine Bourget and Kazunobu Maruyoshi for very useful discussions. The work of A.P~is funded by the Asturian government's SEVERO OCHOA grant BP14-003. A.H. is supported by STFC Consolidated Grant ST/J0003533/1, and EPSRC Programme GrantEP/K034456/1.

\begin{appendices}
 
\numberwithin{equation}{section}

\section{Notation}
\label{app:A}

In all the paper we employed the following conventions
\begin{itemize}
\item  Following \cite{Benvenuti:2010pq} we denote with $t$ the fugacity related to the R-charge of the operators.
\item We denote with the Greek letters $\mu_{i}$, $\nu_{i}$ and $q$ the fugacities of the highest weights of the HWG. More specifically $\nu$ is related to an $SU(2)$ highest weight, while $q$ is related to an $U(1)$ highest weight. 
\item We denote with $x$ the fugacity of the $SU(2)$ global symmetry and with $y_i$ the fugacities of the various $SU(2N)$ or $SO(N)$ global symmetry groups.
\end{itemize}

\section{Details of the computations}
\label{app:quiver}
In this appendix we collect the results for higher values of the number $N$ that parametrizes the quiver gauge theories.
\subsection{N=3, $G_{global}=E_6\supset SU(2) \times SU(6)$}
The fundamental ``building block'' of the quiver diagram reported in figure \ref{fig:e6} is the quiver $[3]-(2)-(1)$. Using the formula (\ref{eq:hsp}) the corresponding Hilbert Series reads
\begin{equation}
\textrm{HS}[T_{(1,1,1)}(SU(3))](t;x_{1}^{(i)},x_{2}^{(i)},x_{3}^{(i)},n_{1},n_{2},0)=t^{2n_{1}}(1-t^{2})^{3}\textrm{PE}\left[t^{2}\sum_{i=1}^{3}\sum_{j=1}^{3}\frac{x_{i}}{x_{j}}\right]\Psi_{U(3)}^{(n_{1},n_{2},0)}(x_{1},x_{2},x_{3};t).
\end{equation}
We glue together three of these theories gauging the common $SU(3)$ flavour group using the formula (\ref{hsglue}). This way the HS for the quiver theory reported in figure \ref{fig:e6} reads
\begin{equation}
\label{eq:hse6}
\begin{split}
& \textrm{HS}_{E_6}(t,\textbf{x}^{(1)},\textbf{x}^{(2)},\textbf{x}^{(3)}) = \sum_{n_{1} \geq n_{2} \geq n_{3}=0} t^{-2n_{1}}(1-t^2)P_{U(3)}(n_{1},n_{2},0;t) \times  \prod_{i=1}^{3}\textrm{HS}[T_{(1,1,1)}(SU(3))](t;\textbf{x}^{(i)},n_{1},n_{2}),
\end{split}\end{equation}
where we set $n_{3}=0$ and the fugacities $\textbf{x}^{i}=(x_{1}^{i},x_{2}^{i},x_{3}^{i})$ satisfy the constraint
\begin{equation}
\prod_{k=1}^{3} x_{k}^{i} = 1, \ \ \textrm{for} \ \  i=1,2,3 \ .
\end{equation}
The corresponding unrefined Hilbert series reads \footnote{This computation has been performed already in \cite{Benvenuti:2010pq}.}
\begin{equation}
\label{hsune6}
\textrm{HS}_{SU(2) \times SU(6)}(t) = \frac{P_{SU(2) \times SU(6)}(t)}{(1-t^2)^{22}},
\end{equation}
where $P_{SU(2)\times SU(6)}(t)$ is a palindromic polynomial given by
\begin{equation}
P_{SU(2) \times SU(6)}(t) = (1+t^2)(1 + 55t^2 +890t^4 + 5886t^6 + 17929t^8 + 26060t^{10} + \ ... \ (\textrm{palindrome}) \ ... + t^{20}). 
\end{equation}
The dimension of the pole of the unrefined HWG (\ref{eq:hwge6}) at $t=1$ is 6 while the degree of the polynomial arising from the dimension of the $SU(2)\times SU(6)$ representation $[n_1+2n_2;n_3,n_4,n_1+2n_5,n_4,n_3]$ is 16. They add up to 22 which is the dimension of the reduced moduli space of one instanton of $E_{6}$.

\subsection{$N=4$, $G_{global} = E_7 \supset SU(8)$}
The fundamental ``building blocks" of the quiver diagram reported in figure \ref{fig:e71} are the quiver diagram [4]-(3)-(2)-(1) and the quiver diagram [4]-(2). We perform the computation of the corresponding Hilbert Series   using the formula (\ref{eq:hsp}). For the first quiver we get 
\begin{equation}\begin{split}
& \textrm{HS}[T_{(1,1,1,1)}(SU(4))](x_{1},x_{2},x_{3},x_{4},n_{1},n_{2},n_{3},0) = \\
&  = t^{(3n_{1}+n_{2}-n_{3})}(1-t^2)^{4}\textrm{PE}\left[\sum_{i=1}^{4}\sum_{j=1}^{4}\frac{x_{i}}{x_{j}}t^2\right]\Psi_{U(4)}^{(n_{1},n_{2},n_{3},0)}(x_{1},x_{2},x_{3},x_{4};t),
\end{split}\end{equation}
where the fugacities $x_{i}$ satisfy the constraint $x_{1}x_{2}x_{3}x_{4}=1$ and we set $n_{4}=0$. While for the second kind of quiver diagram we get
\begin{equation}\begin{split}
& = \textrm{HS}[T_{(2,2)}(SU(4))](p_{1},p_{2},n_{1},n_{2},n_{3},0) = \\ & t^{(3n_{1}+n_{2}-n_{3})}(1-t^2)^{4}\textrm{PE}\left[(2+p_{1}p_{2}^{-1}+p_{2}p_{1}^{-1})(t^{4}+t^2)\right]\Psi_{U(4)}^{(n_{1},n_{2},n_{3},0)}(p_{1}t,p_{1}t^{-1},p_{2}t,p_{2}t^{-1};t),
\end{split}\end{equation}
where the fugacities satisfy the constraint $p_{1}^{2}p_{2}^{2}=1$ and we set $n_{4}=0$.
We use the formula (\ref{hsglue}) and we glue together the three quiver gauge theories gauging the common $SU(4)$ flavour group. The HS of the full quiver gauge theory reads
\begin{equation}
\label{eq:hse7}
\begin{split}
& \textrm{HS}_{E_7}(t;x_1,x_2,x_3,x_4,y_1,y_2,y_3,y_4,p_1,p_2) = \sum_{n_{1} \geq n_{2} \geq n_{3} \geq n_{4}=0} t^{-2(3n_{1}+n_{2}-n_{3})}(1-t^2)P_{U(4)}(n_{1},n_{2},n_{3},0;t)\times \\
& \textrm{HS}[T_{(1,1,1,1)}(SU(4))](t;x_1,x_2,x_3,x_4,n_{1},n_{2},n_{3}) \times \textrm{HS}[T_{(1,1,1,1)}(SU(4))](t;y_1,y_2,y_3,y_4,n_{1},n_{2},n_{3})\times \\
& \textrm{HS}[T_{(2,2)}(SU(4))](p_1,p_2,n_{1},n_{2},n_{3},0).
\end{split}\end{equation}
The unrefined Hilbert series reads\footnote{This result agrees with the expression of the unrefined Hilbert Series of the moduli space of 1-$E_7$ instanton previously found in \cite{Benvenuti:2010pq}.}
\begin{equation}
\label{eq:hsusu8}
\textrm{HS}_{SU(8)}(t) = \frac{P_{SU(8)}(t)}{(1-t^2)^{34}},
\end{equation}
where $P_{SU(8)}(t)$ is a palindromic polynomial given by
\begin{equation}\begin{split}
 P_{SU(8)}(t)  =& 1 + 99 t^{2}+3410 t^{4}+56617 t^{6}+521917 t^{8}+2889898 
   t^{10}+10086066 t^{12}+\\
&   22867856 t^{14} + 34289476 t^{16}+ \ ... \ (\textrm{palindrome}) \ ... \ + t^{34}.
\end{split}\end{equation}
The dimension of the pole of the unrefined HWG (\ref{eq:hwgsu8}) at $t=1$ is 6 while the degree of the polynomial arising from the dimension of the $SU(8)$ representation $[n_1,n_2,n_3,n_4+n_5,n_3,n_2,n_1]$ is 28. They add up to 34 which is the dimension of the reduced moduli space of one-instanton of $E_7$.

\subsection{$N=4$, $G_{global} = SU(2) \times SU(8)$}
The fundamental ``building blocks'' of the quiver diagram reported in fig.\ref{fig:su2su8}  are  the quiver diagram [4]-(3)-(2)-(1) and the quiver diagram [4]-(2)-(1). We perform the computation of the corresponding HS using the formula (\ref{eq:hsp}). The Hilbert Series of the first quiver reads
\begin{equation}\begin{split}
& \textrm{HS}[T_{(1,1,1,1)}(SU(4))](x_{1},x_{2},x_{3},x_{4},n_{1},n_{2},n_{3},0) = \\
&  = t^{3n_{1}+n_{2}-n_{3}}(1-t^2)^{4}\textrm{PE}\left[\sum_{i=1}^{4}\sum_{j=1}^{4}\frac{x_{i}}{x_{j}}t^2\right]\Psi_{U(4)}^{(n_{1},n_{2},n_{3},0)}(x_{1},x_{2},x_{3},x_{4};t),
\end{split}\end{equation}
where the fugacities $x_{i}$ satisfy the constraint $x_{1}x_{2}x_{3}x_{4}=1$ and we set $n_{4}=0$.
While the HS of the second quiver reads
\begin{equation*}\begin{split}
& \textrm{HS}[T_{(2,1,1)}(SU(4))](p_{1},p_{2},p_{3},n_{1},n_{2},n_{3},0) = \\ 
& = t^{3n_{1}+n_{2}-n_{3}}(1-t^2)^{4}K_{(2,1,1)}^{U(4)}(p_{1},p_{2},p_{3},t)\Psi_{U(4)}^{(n_{1},n_{2},n_{3},0)}(p_{1}t,p_{1}t^{-1},p_{2},p_{3};t),
\end{split}\end{equation*}
where  the fugacities satisfy the constraint $p_{1}^{2}p_{2}p_{3}=1$ and we set $n_{4}=0$.
We glue together the three quiver diagrams gauging the common $SU(4)$ global symmetry group using the formula (\ref{hsglue}). 
The corresponding unrefined Hilbert series reads
\begin{equation}
\label{eq:hsu2su8un}
\textrm{HS}_{SU(2) \times SU(8)}(t,1,...,1) = \frac{P_{SU(2) \times SU(8)}(t)}{(1-t)^{36} (1+t)^{24} \left(1+t+t^2 \right)^{18}},
\end{equation}
where $P_{SU(2) \times SU(8)}(t)$ is a palindromic polynomial given by
\begin{equation}
\label{hssu8un}
\begin{split}
& P_{SU(2) \times SU(8)}(t) = 1 +6 t +63 t^{2}+430 t^{3}+2579 t^{4}+13672 t^{5}+64581 t^{6}+273874 t^{7}+1057876 t^{8}+ \\
& 3739708 t^{9}+12168151 t^{10}+36629984 t^{11}+102449081 t^{12}+267099092
   t^{13}+651158236 t^{14}+\\
 &  1488399930 t^{15}+3197185885 t^{16}+6467034500 t^{17}+12340071356 t^{18}+22247726312 t^{19}+\\
& 37949176435 t^{20}+61318520286 t^{21}+93953394952
   t^{22}+136633528532 t^{23}+188739697078 t^{24}+ \\
&  247809311486 t^{25}+309426670826 t^{26}+367597436878 t^{27}+415631203373 t^{28}+447372202126 t^{29}+ \\
& 458475487710 t^{30}+  \ ...  \ + (\textrm{palindrome}) + \ ... \ + \ t^{60}. 
 \end{split}
\end{equation}
The dimension of the pole of the unrefined HWG (\ref{eq:hwgsu2su8new}) at $t=1$ is 7 while the degree of the polynomial arising from the dimension of the $SU(2) \times SU(8)$ representation $[2n_1+n_3;n_2,n_3,n_4,n_5,2n_6+n_3,n_5,n_4,n_2]$ is 29. They add up to 36 which is the dimension of the pole at $t=1$ of the unrefined Hilbert Series (\ref{eq:hsu2su8un}).

\subsection{$N=5$, $G_{global} = SU(10)$}
The ``building blocks'' of the quiver diagram reported in figure \ref{fig:su10} are the quiver diagram [5]-(4)-(3)-(2)-(1) and the quiver diagram [5]-(2). We use the formula (\ref{eq:hsp}) and we  compute the Hilbert Series for each of them.
The Hilbert Series of the first quiver diagram reads
\begin{equation}\begin{split}
& \textrm{HS}[\textrm{T}_{(1,1,1,1,1)}(SU(5))](x_{1},x_{2},x_{3},x_{4},x_{5},n_{1},n_{2},n_{3},n_{4},0) = \\
&  = t^{4n_{1}+2n_{2}-2n_{4}}(1-t^2)^{5}\textrm{PE}\left[\sum_{i=1}^{5}\sum_{j=1}^{5}\frac{x_{i}}{x_{j}}t^2\right]\Psi_{U(5)}^{(n_{1},n_{2},n_{3},n_{4},0)}(x_{1},x_{2},x_{3},x_{4},x_{5};t),
\end{split}\end{equation}
where the fugacities $x_{i}$ satisfy the constraint $x_{1}x_{2}x_{3}x_{4}x_{5}=1$ and we set $n_{5}=0$. The Hilbert Series of the second quiver reads
\begin{equation*}\begin{split}
& \textrm{HS}[\textrm{T}_{(3,2)}[(SU(5))](p_{1},p_{2},n_{1},n_{2},n_{3},n_{4},0) = \\ & t^{4n_{1}+2n_{2}-2n_{4}}(1-t^2)^{5}K_{(3,2)}^{U(5)}(p_{1},p_{2},t)\Psi_{U(5)}^{(n_{1},n_{2},n_{3},n_{4},0)}(p_{1}t^2,p_{1},p_{1}t^{-2},p_{2}t,p_{2}t^{-1};t),
\end{split}\end{equation*}
where the fugacities satisfy the constraint $p_{1}^{3}p_{2}^{2}=1$ and we set $n_{5}=0$.
We glue together the three quiver diagrams gauging the common $SU(5)$ global symmetry group using the formula (\ref{hsglue}).
The corresponding unrefined Hilbert Series reads
\begin{equation}
\label{eq:10exp}
\textrm{HS}_{SU(10)}(t,1,...,1) = \frac{P_{SU(10)}(t)}{(1-t)^{52} (1+t)^{32} \left(1+t+t^2\right)^{26}},
\end{equation}
where $P_{SU(10)}(t)$ is a palindromic polynomial given by 
\begin{equation}
\label{eq:hsu10un}
\begin{split}
& P_{SU(10)}(t) = 1+6 t+88 t^{2}+684 t^{3}+5068 t^{4}+33270 t^{5}+195032 t^{6}+1042038 t^{7}+5115964 t^{8}+\\
& 23174044 t^{9}+97503058 t^{10}+382784562 t^{11}+1407298803 t^{12}+4861817820
   t^{13}+15830225132 t^{14}+\\
& 48702053512 t^{15}+141895604363 t^{16}+392320199370 t^{17}+1031217756368 t^{18}+2581111810032 t^{19}+\\
&  6161027223819 t^{20}+14043326715580
   t^{21}+30604355641425 t^{22}+63836857086540 t^{23}+ \\
&    127576383100320 t^{24}+
   244498094132778 t^{25}+449725582642239 t^{26}+794538443153332 t^{27}+\\
&  1349198124556557 t^{28}+2203437309257322
   t^{29}+3462860983386664 t^{30}+5239624609952376 t^{31}+ \\
&  7636510804862128 t^{32}+10725001017564682 t^{33}+14519955240446539 t^{34}+18955608867282408 t^{35}+\\
& 23869071305501125
   t^{36}+28997597428531974 t^{37}+33994115765048473 t^{38}+38461909792331794 t^{39}+\\
&  42004659363999571 t^{40}+44283517877832144 t^{41}+45070023311322202 t^{42}+44283517877832144 t^{43} +\\
& ... \ \textrm{palindrome} \ ... + t^{84}.
\end{split}\end{equation}
The dimension of the pole of the unrefined HWG (\ref{eq:hwgsu10}) at $t=1$ is 7 while the degree of the polynomial arising from the dimension of the $SU(10)$ representation $[n_1,n_3,n_5,n_6,n_2+n_4,n_6,n_5,n_3,n_1]$ is 45. They add up to 52 which is the dimension of the pole at $t=1$ of the unrefined Hilbert Series (\ref{eq:hsu10un}).

\subsection{$N=5$, $G_{global} = SU(2)\times SU(10)$}
The ``building blocks'' of the quiver diagram reported in figure \ref{fig:su10su2} are the quiver diagram [5]-(4)-(3)-(2)-(1) and the quiver diagram [5]-(2)-(1). We use the formula (\ref{eq:hsp}) and we compute the Hilbert Series of the first quiver diagram 
\begin{equation}\begin{split}
& \textrm{HS}[\textrm{T}_{(1,1,1,1,1)}[(SU(5))](x_{1},x_{2},x_{3},x_{4},x_{5},n_{1},n_{2},n_{3},n_{4},0) = \\
&  = t^{4n_{1}+2n_{2}-2n_{4}}(1-t^2)^{5}\textrm{PE}\left[\sum_{i=1}^{5}\sum_{j=1}^{5}\frac{x_{i}}{x_{j}}t^2\right]\Psi_{U(5)}^{(n_{1},n_{2},n_{3},n_{4},0)}(x_{1},x_{2},x_{3},x_{4},x_{5};t),
\end{split}\end{equation}
where the fugacities $x_{i}$ satisfy the constraint $x_{1}x_{2}x_{3}x_{4}x_{5}=1$ and we set $n_{5}=0$.
The Hilbert Series of the second quiver reads
\begin{equation*}\begin{split}
& \textrm{HS}[\textrm{T}_{(3,1,1)}(SU(5))](p_{1},p_{2},p_{3},n_{1},n_{2},n_{3},n_{4},0) = \\ & = t^{4n_{1}+2n_{2}-2n_{4}}(1-t^2)^{5}K_{(3,1,1)}^{U(5)}(p_{1},p_{2},p_3,t)\Psi_{U(5)}^{(n_{1},n_{2},n_{3},n_{4},0)}(p_{1}t^2,p_{1},p_{1}t^{-2},p_{2},p_{3};t),
\end{split}\end{equation*}
where this time the fugacities satisfy the constraint $p_{1}^{3}p_{2}p_{3}=1$ and we set $n_{5}=0$.
We use the formula (\ref{hsglue}) and  we compute the Hilbert Series of the quiver reported in figure \ref{fig:su10su2}. 
The corresponding unrefined Hilbert series  reads
\begin{equation}
\label{eq:hsunsu2su10}
\textrm{HS}_{SU(2) \times SU(10)}(t,1,...,1) = \frac{P_{SU(2) \times SU(10)}(t)}{\left(1-t^2\right)^{54} \left(1+t^2\right)^{27}},
\end{equation}
where $P_{SU(2) \times SU(10)}(t)$ is a palindromic polynomial given by 
\begin{equation}\begin{split}
& P_{SU(2) \times SU(10)}(t) = 1 +75 t^{2}+3227 t^{4}+93628 t^{6}+1995005 t^{8}+32743316 t^{10}+428018495 t^{12}+ \\
& 4567431397 t^{14}+40562120142 t^{16}+304454872921 t^{18}+1956103026075
   t^{20}+10872201648590 t^{22}+ \\
& 52743372310579 t^{24}+225023527079799 t^{26}+849778417559022 t^{28}+2856352728077173 t^{30}+\\
& 8586607150389539 t^{32}+23180525972202894
   t^{34}+56397132911896665 t^{36}+124035385935159636 t^{38}+\\
&  247240406636319041 t^{40}+447648087747618549 t^{42}+737563815297879763 t^{44}+1107559743906553834
   t^{46}+\\
&   1517629622095781174 t^{48}+1899315502205824974 t^{50}+2172420593881086252 t^{52}+2271821768928281868 t^{54} +\\
& \ ... \textrm{palindrome} + \ ... \ + t^{108}.
\end{split}\end{equation}
We observe that the dimension of the pole of the unrefined HWG (\ref{eq:hwgsu2su10}) at $t=1$ is 8 while the degree of the polynomial arising from the dimension of the $SU(2) \times SU(10)$ representation $[2n_2+n_4;n_1,n_3,n_5,n_6,n_4+2n_7,n_6,n_5,n_3,n_1]$ is 46. They add up to 54 which is the dimension of the pole at $t=1$ of the unrefined Hilbert Series.

\subsection{$N=6$, $G_{global}=SU(12)$}
The ``building blocks'' of the quiver diagram reported in fig.\ref{fig:su12} are the quiver diagram [6]-(5)-(4)-(3)-(2)-(1) and the quiver diagram [6]-(2). We use the formula (\ref{eq:hsp}) and we compute the corresponding Hilbert Series.  The HS of the first quiver reads
\begin{equation}\begin{split}
& \textrm{HS}[\textrm{T}_{(1,1,1,1,1,1)}(SU(6))](x_{1},x_{2},x_{3},x_{4},x_{5},x_6,n_{1},n_{2},n_{3},n_{4},n_5,0) = \\
&  = t^{5n_1+3n_2+n_3-n_4-3n_5-5n_6}(1-t^2)^{6}\textrm{PE}\left[\sum_{i=1}^{6}\sum_{j=1}^{6}\frac{x_{i}}{x_{j}}t^2\right]\Psi_{U(6)}^{(n_{1},n_{2},n_{3},n_{4},n_5,0)}(x_{1},x_{2},x_{3},x_{4},x_{5},x_{6};t),
\end{split}\end{equation}
where the fugacities $x_{i}$ satisfy the constraint $x_{1}x_{2}x_{3}x_{4}x_{5}x_6=1$ and we set $n_{6}=0$.
The Hilbert Series of the second quiver reads
\begin{equation*}\begin{split}
& \textrm{HS}[\textrm{T}_{(4,2)}[(SU(6))](p_{1},p_{2},p_{3},n_{1},n_{2},n_{3},n_{4},n_5,0) = t^{5n_1+3n_2+n_3-n_4-3n_5-5n_6}(1-t^2)^{6}\\
& K_{(4,2)}^{U(6)}(p_{1},p_{2},t)\Psi_{U(6)}^{(n_{1},n_{2},n_{3},n_{4},n_{5},0)}(p_{1}t^{3},p_{1}t^{1},p_{1}t^{-1},p_{1}t^{-3},p_{2}t^{1},p_{2}t^{-1};t),
\end{split}\end{equation*}
where  the fugacities satisfy the constraint $p_{1}^{4}p_{2}^2=1$ and we set $n_{6}=0$.

\subsection{$N=6$, $G_{global}=SU(2) \times SU(12) $}
The ``building blocks'' of the quiver diagram reported in fig.\ref{fig:su12su2}  are the quiver diagram [6]-(5)-(4)-(3)-(2)-(1) and the quiver diagram [6]-(2). We use the formula (\ref{eq:hsp}) and we find the Hilbert Series for each of them. The Hilbert Series of the first quiver diagram reads
\begin{equation}\begin{split}
& \textrm{HS}[\textrm{T}_{(1,1,1,1,1,1)}(SU(6))](x_{1},x_{2},x_{3},x_{4},x_{5},x_6,n_{1},n_{2},n_{3},n_{4},n_5,0) = \\
&  = t^{5n_1+3n_2+n_3-n_4-3n_5-5n_6}(1-t^2)^{6}\textrm{PE}\left[\sum_{i=1}^{6}\sum_{j=1}^{6}\frac{x_{i}}{x_{j}}t^2\right]\Psi_{U(6)}^{(n_{1},n_{2},n_{3},n_{4},n_5,0)}(x_{1},x_{2},x_{3},x_{4},x_{5},x_{6};t),
\end{split}\end{equation}
where the fugacities $x_{i}$ satisfy the constraint $x_{1}x_{2}x_{3}x_{4}x_{5}x_6=1$ and we set $n_{6}=0$.
The HS of the second quiver diagram is
\begin{equation*}\begin{split}
& \textrm{HS}[\textrm{T}_{(4,1,1)}(SU(6))](p_{1},p_{2},p_{3},n_{1},n_{2},n_{3},n_{4},n_5,0) = t^{5n_1+3n_2+n_3-n_4-3n_5-5n_6}(1-t^2)^{6}\\
& K_{(4,1,1)}^{U(6)}(p_{1},p_{2},p_3,t)\Psi_{U(6)}^{(n_{1},n_{2},n_{3},n_{4},n_{5},0)}(p_{1}t^{3},p_{1}t,p_{1}t^{-1},p_{1}t^{-3},p_{2},p_{3};t),
\end{split}\end{equation*}
where the fugacities satisfy the constraint $p_{1}^{4}p_{2}p_3=1$ and we set $n_{6}=0$.
We glue together the Hilbert Series of the three quivers diagrams using the formula (\ref{hsglue}).

\subsection{$G_{global}=SO(14) \times U(1)$}
The quiver diagram with $SO(14)\times U(1)$ global symmetry is reported in figure \ref{fig:so14u1}.

The ``building blocks'' of this theory are the quiver diagram [5]-(4)-(3)-(2)-(1) and the quiver diagram [5]-(3)-(1). We use the formula (\ref{eq:hsp}) and we find the HS of each ``building blocks''. The HS of the quiver [5]-(4)-(3)-(2)-(1) reads
\begin{equation}\begin{split}
& \textrm{HS}[\textrm{T}_{(1,1,1,1,1)}(SU(5))](x_{1},x_{2},x_{3},x_{4},x_{5},n_{1},n_{2},n_{3},n_{4},0) = \\
&  = t^{4n_{1}+2n_{2}-2n_{4}}(1-t^2)^{5}\textrm{PE}\left[\sum_{i=1}^{5}\sum_{j=1}^{5}\frac{x_{i}}{x_{j}}t^2\right]\Psi_{U(5)}^{(n_{1},n_{2},n_{3},n_{4},0)}(x_{1},x_{2},x_{3},x_{4},x_{5};t),
\end{split}\end{equation}
where the fugacities $x_{i}$ satisfy the constraint $x_{1}x_{2}x_{3}x_{4}x_{5}=1$ and we set $n_{5}=0$.
The HS of the quiver [5]-(3)-(1) reads 
\begin{equation}\begin{split}
& \textrm{HS}[\textrm{T}_{(2,2,1)}(SU(5))](y_{1},y_{2},y_{3},n_{1},n_{2},n_{3},n_{4},0) = \\
&  = t^{4n_{1}+2n_{2}-2n_{4}}(1-t^2)^{5}K_{(2,2,1)}^{U(5)}(y_1,y_2,y_3,t)\Psi_{U(5)}^{(n_{1},n_{2},n_{3},n_{4},0)}(y_{1},y_{2},y_{3};t),
\end{split}\end{equation}
where the fugacities $y_{i}$ satisfy the constraint $y_{1}^2y_{2}^2y_{3}=1$ and we set $n_{5}=0$. The factor $K_{(2,2,1)}^{U(5)}$ is given by
\begin{equation*}
K_{(2,2,1)}^{U(5)}(y_1,y_2,y_3,t) = \textrm{PE}\left[ 3t^2 + 2t^4 + \frac{t^2y_1}{y_2} + \frac{t^4 y_1}{y_2} + \frac{t^2 y_2}{y_1} + \frac{t^4y_2}{y_1} +\frac{t^3 y_1}{y_3} + \frac{t^3 y_2}{y_3} + \frac{t^3 y_3}{y_1} + \frac{t^3 y_3}{y_2}\right].
\end{equation*}
We glue together the three gauge theories using the formula (\ref{hsglue}).
The corresponding unrefined HS reads
\begin{equation}
\label{eq:unhss014}
\begin{split}
& \textrm{HS}_{SO(14)\times U(1)}(t) = \\
& = \frac{1}{(1-t)^{44} (1+t)^{36} \left(+1+t+t^2 \right)^{22}}( 1+14 t+161 t^{2}+1450 t^{3}+11235 t^{4}+76076 t^{5}+460270 t^{6}+\\
&  2515464
   t^{7}+12543757 t^{8}+57485438 t^{9}+243590526 t^{10}+959135368 t^{11}+3524162306
   t^{12}+\\
& 12127021450 t^{13}+39204806082 t^{14}+119399888418 t^{15}+343401526770
   t^{16}+934694624360 t^{17}+\\
& 2412360246363 t^{18}+5913823733836 t^{19}+13791838727022
   t^{20}+30641576394730 t^{21}+\\
&   64935808673035 t^{22}+131413027444364 t^{23}+254228078480137
   t^{24}+470594664975578 t^{25}+\\
&  834214522978169 t^{26}+1417259205143370
   t^{27}+2309205745322137 t^{28}+3610677213615316 t^{29}+\\
&   5420885724624754
   t^{30}+7818510084828412 t^{31}+10837817133061051 t^{32}+14444157990254486
   t^{33}+\\   
&  18514905671456643 t^{34}+22832559942436902 t^{35}+27095427129072378
   t^{36}+30947837022133142 t^{37}+\\
&  34026931702773205 t^{38}+36018018650672498
   t^{39}+36707016757599132 t^{40}+ ... + \textrm{palindrome} + t^{80}) \ .
\end{split}\end{equation}
The dimension of the pole arising from the HWG (\ref{eq:hwgso14u1}) is 5 while the degree of the polynomial arising from the $[0,n_2,0,n_4,0,n_6,n_7]$ representation is 39. They add up to 44 which is the dimension of the pole of the unrefined HS (\ref{eq:unhss014}) .

\end{appendices}

\bibliographystyle{ytphys}
\small\baselineskip=.97\baselineskip
\bibliography{ref}

\end{document}